\documentclass[10pt
]{article}
\usepackage[a4paper]{geometry}
\usepackage{graphicx}
\usepackage{amsmath,amssymb}
\usepackage{authblk}
\usepackage{cite}
\usepackage{subcaption}
\usepackage{xfrac}
\usepackage{authblk}

\usepackage{tikz}
\usetikzlibrary{decorations.pathmorphing}
\usetikzlibrary{decorations.markings}
\usetikzlibrary{shadows}
\usepackage{pgfplots}

\usepackage[abs]{overpic}

\usepackage[textsize=small,shadow]{todonotes} 

\usepackage[defaultsans]{droidsans}
\usepackage[eulergreek]{sansmath}
\pgfplotsset{
  tick label style = {font=\sansmath\sffamily},
  every axis label = {font=\sansmath\sffamily},
  legend style = {font=\sansmath\sffamily},
  label style = {font=\sansmath\sffamily}
}

\newcommand{\Rho}{\mathrm{P}}
\renewcommand{\vec}[1]{{\mathbf{#1}}}
\newcommand{\vect}[1]{{\boldsymbol{#1}}}
\newcommand\salto[1]{[\mskip-3mu[ #1 ]\mskip-3mu]}
\DeclareMathOperator{\Hankel}{H}

\title{Transformation elastodynamics and cloaking for flexural waves}

\author[1,2]{D.J. Colquitt}
\author[2,3]{M. Brun\thanks{Author for correspondence: mbrun@unica.it}}
\author[4]{M. Gei}
\author[2]{A.B. Movchan}
\author[2]{N.V. Movchan}
\author[5]{I.S. Jones}

\affil[1]{\small Department of Mathematics, Imperial College London, South Kensington, London, SW7 2AZ, U.K.}
\affil[2]{Department of Mathematical Sciences, University of Liverpool, Liverpool, L69 3BX, U.K.}
\affil[3]{Dipartimento di Ingegneria Meccanica, Chimica e dei Materiali, Universit\'{a} di Cagliari, Piazza d'Armi, I-09123 Cagliari, Italy}
\affil[4]{Department of Civil, Environmental and Mechanical Engineering, University of Trento, I-38123 Trento, Italy}
\affil[5]{School of Engineering, John Moores University, Liverpool, L3 3AF, U.K.}

\date{June 9, 2014}

\begin{document}

\maketitle

\begin{abstract}
The paper addresses an important issue of cloaking transformations for fourth-order partial differential equations representing flexural waves in thin elastic plates.
It is shown that, in contrast with the Helmholtz equation, the general form of the partial differential equation is not invariant with respect to the cloaking transformation.
The significant result of this paper is the analysis of the transformed equation and its interpretation in the framework of the linear theory of pre-stressed plates.
The paper provides a formal framework for transformation elastodynamics as applied to elastic plates.
Furthermore, an algorithm is proposed for designing a broadband square cloak for flexural waves, which employs a regularised push-out transformation.
Illustrative numerical examples show high accuracy and efficiency of the proposed cloaking algorithm. In particular, a physical configuration involving a perturbation of an interference pattern generated by two coherent sources is presented. It is demonstrated that the perturbation produced by a cloaked defect is negligibly small even for such a delicate interference pattern.
\end{abstract}

\begin{center}
\end{center}

\section{Introduction}

Transformation optics, as introduced by Leonhardt~\cite{leonhardt2006} and Pendry~\cite{pendry2006}, has proven to be a useful tool in the design and fabrication of invisibility cloaks for electromagnetic waves (see~\cite{schurig2006,landy2012,ergin2011,chen2012}, among many others).
Prior to cloaking, the ideas of transformation optics were used as a computational tool~\cite{ward1996} to aid in simulations involving several length scales or complex geometries.
The cornerstone of transformation optics is the invariance of Maxwell's equations under coordinate transformations~\cite{post1962}.
In essence, the design of invisibility cloaks via transformation optics involves deforming a region of space such that a disc is mapped to an annulus.
Electromagnetic waves then propagate around the annulus as if it were a disc and, in this sense, any object placed inside the inner annulus is said to be invisible.
The ideas behind transformation optics have also been successfully applied to systems that are governed by equations isomorphic to Maxwell's equations, such as acoustics~\cite{chen2007,cummer2007,norris2008,chen2010}, thermodynamics~\cite{guenneau2012thermo,schittny2013,han2013}, and out-of-plane elastic waves~\cite{parnell2012,parnell2013,colquitt2013}.
Metamaterials have also found broad application in a wide range of physical settings~\cite{kadic2013}.

However, in general, the partial differential equations governing physical systems are not invariant under coordinate transformations.
In particular, the elastodynamic wave equation is not invariant under a general mapping~\cite{milton2006,norris2011}.
Norris \& Shuvalov~\cite{norris2011} showed that the transformed equation requires either non-symmetric stress or tensorial density.
However, Milton et al.~\cite{milton2006} demonstrated that, for a particular choice of gauge, a more general constitutive model (the so-called Willis model~\cite{milton2007}) remains invariant under an arbitrary coordinate transformation yielding symmetric stress but tensorial density.
Brun et al.~\cite{brun2009} applied a cloaking transformation to the Navier equations for isotropic linear elasticity and found that the transformed equations correspond to non-symmetric constitutive relations where only major symmetry is preserved.
More recently Norris \& Parnell~\cite{norris2012} showed that it is possible to obtain the constitutive equations required for elastodynamic cloaking by the application of a finite pre-strain.
However, using this approach there is a limit on the relative size of the cloaked region and regularisation parameter used to create the near cloaks.

Transformation elastodynamics has also been used in the design of invisibility cloaks for flexural waves in 
thin elastic plates~\cite{farhat2009PRL,farhat2009} with corresponding experiments recently performed by the group led by Wegener~\cite{stenger2012}.
The equations governing the motion of flexural waves in thin plates are not, in general, invariant under coordinate transformations. Nevertheless, Farhat et al.~\cite{farhat2009PRL} found that the transformed equation can be identified with a thin plate if one assumes strains of the von-K\'arm\'an form (see, for example,~\cite{timoshenko1959,lekhnitskii1968}).
However, as demonstrated by Ciarlet~\cite{ciarlet1980} and Blanchard \& Ciarlet~\cite{blanchard1983} the theory of von-K\'arm\'an corresponds to the leading order behaviour of a thin elastic plate under moderate deformation provided that the applied loads and/or the elastic moduli have a specific dependence on the thickness of the plate.

The present paper is concerned with the construction of a transformation theory for the dynamic equations of flexural deformations in Kirchhoff-Love plates.
In contrast to the elastodynamic case~\cite{milton2006,norris2011,norris2012,brun2009} and previous work on plates~\cite{farhat2009PRL,farhat2009}, it is shown that it is possible to construct an invisibility cloak for flexural waves in thin plates without recourse to non-symmetric stresses, tensorial densities, or non-linear theories.
In particular, it is shown that by the application of appropriate in-plane forces it is possible to construct an exact invisibility cloak for flexural waves within the framework of linear Kirchhoff-Love plate theory.
This result could lead to a refinement of the experimental implementation of invisibility cloaks for flexural waves yielding an improvement of the experimental results reported in~\cite{stenger2012}.

The structure of the paper is as follows.
After the introduction, a formal framework for transformation elastodynamics applied to Kirchhoff-Love plates is developed.
The material parameters in the transformed system are given explicitly in terms of the deformation gradient.
In \S\ref{sec:trans-plates}, a general coordinate transformation is applied to the equation, which governs the flexural displacement of a Kirchhoff-Love plate.
It is demonstrated that, in general, the bi-harmonic operator is not invariant under a cloaking transformation.
However in \S\ref{sec:generalised-gov-eqn}, it is shown that a sensible physical interpretation can be given to the transformed equation corresponding to a pre-stressed linear anisotropic inhomogeneous Kirchhoff-Love plate.
The natural and essential interface conditions are discussed in \S\ref{sec:interface}. A regularised cloaking push-out transformation \cite{colquitt2013} is discussed in \S\ref{sec:example-cloak}.
The material parameters of the cloak and the applied pre-stresses are given explicitly in \S\ref{sec:mat-props}, where it is shown that it is possible to reduce the fully anisotropic plate to a locally orthortropic plate.
A series of illustrative simulations are presented in \S\ref{sec:sims}, along with a comparison with the corresponding problem for the membrane in \S\ref{sec:corresponding-helmholtz}.
The efficiency of the cloak is examined in \S\ref{sec:quality} while the quality of the cloaking effect is demonstrated using a delicate interference pattern in \S\ref{sec:interference}.
The paper concludes with some remarks in \S\ref{sec:conclusion}.

\section{Transforming Kirchhoff-Love plates}
\label{sec:trans-plates}

In the absence of applied in-plane forces, the equation governing the time-harmonic out-of-plane displacement amplitude $w(\vec{X})$ of an isotropic homogeneous Kirchhoff-Love plate under pure bending is~\cite{timoshenko1959,lekhnitskii1968}
\begin{equation}
\left(\nabla_\vec{X}^4 -  \frac{\Rho h}{D^{(0)}}\omega^2\right)w(\vec{X}) = 0,\;\vec{X}\in\chi\subseteq\mathbb{R}^2
\label{eq:untransformed-plate-eq}
\end{equation}
where $D^{(0)}$, $\Rho$, and $h$ are the flexural rigidity, density, and thickness of the plate respectively; and $\omega$ is the radian frequency.
Consider an invertible transformation $\mathcal{F}:\chi\mapsto\Omega$ and $\vect{x} = \mathcal{F}(\vec{X})$.
By a double application of~\cite[Lemma 2.1]{norris2008} equation~\eqref{eq:untransformed-plate-eq} in new coordinates may be expressed as
\begin{equation}
\left(\nabla\cdot J^{-1}\vec{F}\vec{F}^\mathrm{T}\nabla J \nabla\cdot J^{-1}\vec{F}\vec{F}^\mathrm{T}\nabla - \frac{\Rho h}{JD^{(0)}}\omega^2\right)w(\vect{x}) = 0,\;\vect{x}\in\Omega,
\label{eq:transformed-plate-eq}
\end{equation}
where $\nabla = \nabla_\vect{x}$, $\vec{F} = \nabla_\vec{X}\vect{x}$  is the deformation gradient and $J = \det\vec{F}$ is the Jacobian.
For ease of exposition in what follows it is convenient to work in Cartesian coordinates. The differentiation in the equations below is applied with respect to the vector 
variable $\vect{x}$ in the transformed domain. 
Using the Einstein summation convention the transformed equation~\eqref{eq:transformed-plate-eq} expressed in Cartesian coordinates is
\begin{multline}
JG_{ij}G_{kl}\,w_{,ijkl} + 2 \left(JG_{ij}G_{kl}\right)_{,i}w_{,jkl} \\
+ \left[ G_{ij}(JG_{k l})_{,ij} + 2G_{jk}\left(JG_{il,i}\right)_{,j} + G_{ij,i}(JG_{k l})_{,j} + JG_{ik,i}G_{jl,j}\right]w_{,k l} \\
+\left[G_{ij,i}\left(JG_{k l,k}\right)_{,j} +G_{ij}\left(JG_{k l,k}\right)_{,ij}\right]w_{,l}
 - \frac{\Rho h}{JD^{(0)}}\omega^2w = 0,
\label{eq:transformed-plate-eq-index}
\end{multline}
where the symmetric tensor 
\begin{equation}
G_{ij} = J^{-1} F_{ip}F_{jp}
\label{eq:Gij}
\end{equation}
has been introduced and subscript commas followed by indices indicate differentiation with respect to spatial variables.
It is emphasised that the governing equation for flexural vibrations in a thin elastic plate is not invariant under coordinate transformation as can be seen from \eqref{eq:transformed-plate-eq-index}.

Hence, there is a difference, compared to a similar procedure applied to the transformation of the Helmholtz equation, as used in models of invisibility cloaks in acoustics or electromagnetism (see, for example,~\cite{leonhardt2006,pendry2006,cummer2007,norris2008}).
In particular, the transformed Helmholtz equation, with the appropriate choice of the gauge, can be interpreted as the governing equation for time-harmonic waves  in an inhomogeneous anisotropic medium. There was no need for the introduction of any additional external fields in such models.   
In contrast, equation~\eqref{eq:transformed-plate-eq-index} is not the standard form for an anisotropic inhomogeneous plate. Namely, as in \cite{lekhnitskii1968}, the equation of anisotropic inhomogeneous plate, of flexural rigidities  $D_{ijkl}(\vect{x})$, has the form  
\begin{equation}
D_{ijkl}w_{,ijkl} + 2D_{ijkl,i}w_{,jkl} + D_{ijkl,ij}  w_{,kl}   - h\rho\omega^2 w =0 .
\label{eq:eom1}
\end{equation}
Whereas the first two terms in \eqref{eq:eom1} have the same structure as in \eqref{eq:transformed-plate-eq-index}, there is a discrepancy in the structure of the remaining terms involving first and second-order derivatives of $w$ in the above equations  \eqref{eq:transformed-plate-eq-index}  and \eqref{eq:eom1}. This discrepancy cannot be rectified by any choice of elastic constants or inertia in the transformed domain. 

In this paper, it is shown that a physical interpretation can be given to the transformed equation \eqref{eq:transformed-plate-eq-index}, subject to the introduction of an appropriate pre-stress.
Furthermore, this approach will be used to construct an  invisibility cloak for flexural waves within the framework of anisotropic pre-stressed plate theory.

As in~\cite{norris2008,norris2011,colquitt2013}, the transformations considered in the present paper is assumed to be invertible.
\emph{Perfect} cloaks require that the transformation be singular on the inner boundary of the cloak, which leads to cloaks with singular material properties.
However, a regularisation procedure can lead to construction of \emph{near cloaks} as introduced by Kohn et al.~\cite{kohn2008}.
The regularisation parameter $\epsilon$ can be taken as small as desired in order to achieve the required accuracy of the cloak.

\subsection{Governing equations in the presence of in-plane forces}
\label{sec:generalised-gov-eqn}

The following discussion provides the physical interpretation of equation~\eqref{eq:transformed-plate-eq-index} in the transformed domain. 
In the presence of in-plane forces, the time-harmonic flexural deformation of a Kirchhoff-Love plate is governed by the following equation (see \cite{timoshenko1959,lekhnitskii1968,leissa1969})
\begin{equation}
M_{ij,ij} + N_{ij}w_{,ij} - S_{i}w_{,i}  = -h\rho\omega^2 w,
\end{equation}
where $w$ is the out-of-plane displacement.
Here the plate is  
subjected to membrane forces ($N_{ij}$) and in-plane body forces ($S_i$). 
Additionally, the membrane and in-plane body forces are constrained to satisfy the equilibrium equation
\begin{equation}
N_{ij,j} + S_i = 0.
\label{eq.membrane}
\end{equation}
Using the constitutive equation for a linearly elastic Kirchhoff-Love plate, $M_{ij} = -D_{ijkl}(\vect{x})w_{,kl}$, the equation of motion becomes
\begin{equation}
D_{ijkl}w_{,ijkl} + 2D_{ijkl,i}w_{,jkl} + \left(D_{ijkl,ij} - N_{kl}\right)w_{,kl} + S_{l}w_{,l}  = h\rho\omega^2 w,
\label{eq:eom}
\end{equation}
where $D_{ijkl}$ are the flexural rigidities.
Using equation~\eqref{eq:eom}, the terms in the transformed equation~\eqref{eq:transformed-plate-eq-index} may be identified with physically meaningful quantities.
In this manner, the transformed equation governing the flexural displacement of a Kirchhoff-Love plate under an arbitrary coordinate mapping may be interpreted as a generalised plate.
It is emphasised that the generalised model remains within the framework of linear Kirchhoff-Love theory.
In particular, the flexural rigidities of the transformed plate are immediately identified
\begin{subequations}
\begin{equation}
D_{ijkl} = D^{(0)}JG_{ij}G_{kl},
\label{eq:flexural-rigidities}
\end{equation}
as are the in-plane body forces
\begin{equation}
S_l = D^{(0)}\left[G_{jk}\left(JG_{il,i}\right)_{,j}\right]_{,k},
\label{eq:body-forces}
\end{equation}
and the transformed density
\begin{equation}
\rho = \frac{P}{J}.
\end{equation}
It is clear that the flexural rigidities~\eqref{eq:flexural-rigidities} possess the expected major and minor symmetries.
Hence, in general, there are six independent elastic parameters required to define a platonic cloak.
In addition to the above, the transformed equation~\eqref{eq:transformed-plate-eq-index} should also satisfy two additional constraints.
Firstly, the coefficients of the third order terms in~\eqref{eq:transformed-plate-eq-index} must match the derivatives of the flexural rigidities, that is, $D_{ijkl,i} = D^{(0)}(JG_{ij}G_{kl})_{,i}$.

Secondly, the membrane forces $N_{kl}$ must be chosen such that the second order terms in~\eqref{eq:transformed-plate-eq-index} match those in~\eqref{eq:eom}.
Further, the membrane and in-plane body forces must also satisfy the equilibrium equation \eqref{eq.membrane}.
The membrane forces appear only in terms involving the second order derivative and it is assumed that $w$ is sufficiently smooth to allow the order of differentiation to be interchanged.
These forces are obtained by integrating the stresses through the thickness of the plate
\[
N_{kl} = \int\limits_{-h/2}^{h/2}\sigma_{kl}\,\mathrm{d}z,
\]
whence $N_{kl} = N_{l k}$ is required for symmetric stress.
The desired symmetry is obtained by taking $N_{kl}$ in the form
\begin{equation}
N_{kl} = D^{(0)}\left[\left(JG_{kl}G_{ij,i} - JG_{jl}G_{ik,i}\right)_{,j} - 
G_{jk}\left(JG_{il,i}\right)_{,j}\right].
\end{equation}
It is now straightforward to verify that the membrane and body forces satisfy the in-plane equilibrium equation \eqref{eq.membrane}.
\label{eq:mat-params}
\end{subequations}

It has thus been demonstrated that, under a general coordinate mapping, the equation governing time-harmonic flexural vibrations of a linear isotropic homogeneous Kirchhoff-Love plate transforms to an equation corresponding to a linear anisotropic inhomogeneous Kirchhoff-Love plate
in the presence of in-plane loads.
It is emphasised that these loads depend only on the coordinate mapping (via the deformation gradient) and are not functions of displacement nor time.
In this sense, the membrane forces $N_{kl}$ can be interpreted as a pre-stress together with appropriate body forces $S_{l}$ to ensure equilibrium.
This formalism represents a general framework in which transformation elastodynamics for Kirchhoff-Love plates can be investigated.
The distinguishing feature of this interpretation
is that, although a generalised plate model is introduced, the framework is entirely linear and all terms are identified with well understood physical quantities.

It is observed that compressive pre-stress may occur (see the specific example in \S\ref{sec:mat-props}) and, as a result, the question of buckling may arise in the practical implementation. The buckling load depends
not only on the geometrical and material parameters of the structure, but also on the distribution of the pre-stress arising from the particular geometric transformation implemented.
The maximum compressive stress occurs at the interior boundary of the cloak in neighbourhoods of the corner points and its magnitude 
is governed by the regularisation parameter $\epsilon$ used to create the near cloaks.

A further notable feature of the new framework, as above, is that it ensures that the stiffnesses have both major and minor symmetries, the stresses are symmetric and the transformed density is scalar.
This is in contrast to the case of cloaks for  vector  three- and two-dimensional elasticity~\cite{norris2011} where there is either a non-symmetric stress~\cite{brun2009} or tensorial mass density and dependence of stress on velocity~\cite{milton2006}. It is also appropriate to mention the use of pre-stress in problems of control of elastic waves for anti-plane and in-plane wave motion in hyperelastic materials \cite{parnell2012,parnell2013,norris2012} and interface incremental problems of vector elasticity with finite pre-stress \cite{Bigoni2008,Gei2009}. In the latter case, the emphasis is on the modelling and control of Floquet-Bloch waves in a surface elastic layer subjected to a finite pre-stress and localisation near defects, rather than re-routing of elastic waves around an obstacle.

\subsection{Interface conditions}
\label{sec:interface}

Field equations (\ref{eq:transformed-plate-eq}) or (\ref{eq:eom}) are  
accompanied by 
transmission conditions on the interface boundary between  
transformed and untransformed domains. 
The interface conditions can be deduced by applying the principle of virtual displacement  (see \cite{timoshenko1959} among others) and extending the result  to 
anisotropic media.
The essential 
interface conditions are the continuity of transverse displacement and its normal derivative on the interface between two domains:
\begin{equation}
\salto{w}=0, \quad
\left[\!\!\left[\,\frac{\partial w}{\partial n}\,\right]\!\!\right]=0, 
\quad
 \mbox{on } \partial\Omega,
\label{eq:int-ess}
\end{equation}
where $\salto{.}$ denotes the jump and ${\bf n}$ is the unit normal on the interface $\partial\Omega$.
The natural interface conditions correspond to the continuity of the vertical forces and of the normal component of the bending moment across the domain interface. They can be expressed in term of $w(\bf x)$ in the form
\begin{equation}
\left[\!\!\left[\left(N_{ij}\,w_{,j}-{(D_{ijkl}w_{,kl})}_{,j}-
\frac{\partial}{\partial s}\left(D_{ijkl}w_{,kl}s_j\right)\right)\,n_i\right]\!\!\right]=0, \quad
\left[\!\left[\,D_{ijkl}w_{,kl}\,n_jn_i\right]\!\right]=0, \quad \mbox{on } \partial\Omega,
\label{eq:int-nat}
\end{equation}
where $\vec{s}$ is the anti-clockwise unit tangent on $\partial\Omega$.

\section{A push-out transformation: the square cloak}
\label{sec:example-cloak}

\begin{figure}
\centering
\begin{tikzpicture}[scale=0.7]
\begin{scope}[shift={(-5.25,0)}]
\draw[fill=lightgray] decorate [decoration={random steps,segment length=10,amplitude=5}] {(0,0) circle (5)};
\draw[fill=lightgray] (-2.5,-2.5) rectangle (2.5,2.5);
\draw[fill=white] (-0.5,-0.5) rectangle (0.5,0.5);
\draw (0.5,0.5) -- (2.5,2.5);
\draw (-0.5,0.5) -- (-2.5,2.5);
\draw (-0.5,-0.5) -- (-2.5,-2.5);
\draw (0.5,-0.5) -- (2.5,-2.5);
\node at (0,0) {$\displaystyle \chi^{(0)}$};
\node at (1.5,0) {$\displaystyle \chi^{(1)}$};
\node at (0,1.5) {$\displaystyle \chi^{(2)}$};
\node at (-1.5,0) {$\displaystyle \chi^{(3)}$};
\node at (0,-1.5) {$\displaystyle \chi^{(4)}$};
\node[below] at (0,-4) {$\displaystyle\color{black} \chi_+$};
\node[right] at (2.5,1.5) {$\displaystyle\color{black} \Gamma^{(1)}$};
\node[above] at (-1.5,2.5) {$\displaystyle \color{black}\Gamma^{(2)}$};
\node[left] at (-2.5,-1.5) {$\displaystyle\color{black} \Gamma^{(3)}$};
\node[below] at (1.5,-2.5) {$\displaystyle\color{black} \Gamma^{(4)}$};
\end{scope}
\begin{scope}[shift={(5.25,0)}]
\draw[fill=lightgray] decorate [decoration={random steps,segment length=10,amplitude=5}] {(0,0) circle (5)};
\draw[fill=gray] (-2.5,-2.5) rectangle (2.5,2.5);
\draw[fill=white] (-1.5,-1.5) rectangle (1.5,1.5);
\draw (1.5,1.5) -- (2.5,2.5);
\draw (-1.5,1.5) -- (-2.5,2.5);
\draw (-1.5,-1.5) -- (-2.5,-2.5);
\draw (1.5,-1.5) -- (2.5,-2.5);
\node at (0,0) {$\displaystyle \Omega^{(0)}$};
\node at (2,0) {$\displaystyle\color{white} \Omega^{(1)}$};
\node at (0,2) {$\displaystyle\color{white} \Omega^{(2)}$};
\node at (-2,0) {$\displaystyle\color{white} \Omega^{(3)}$};
\node at (0,-2) {$\displaystyle\color{white} \Omega^{(4)}$};
\node[below] at (0,-4) {$\displaystyle\color{black} \Omega_+$};
\node[right] at (2.5,1.5) {$\displaystyle\color{black} \Gamma^{(1)}$};
\node[above] at (-1.5,2.5) {$\displaystyle \color{black}\Gamma^{(2)}$};
\node[left] at (-2.5,-1.5) {$\displaystyle\color{black} \Gamma^{(3)}$};
\node[below] at (1.5,-2.5) {$\displaystyle\color{black} \Gamma^{(4)}$};
\end{scope}
\draw[->, bend left,line width = 2] (-5,4) to (5,4);
\node[below] at (0,5) {$\displaystyle\mathcal{F}$};
\end{tikzpicture}
\caption{\label{fig:deformation}
The map $\mathcal{F}$ transforms the undeformed region $\chi = \left(\cup_{i=1}^4 \chi^{(i)}\right)\cup \chi_+$ to the deformed configuration $\Omega = \left(\cup_{i=1}^4 \Omega^{(i)}\right)\cup\Omega_+$.
The exterior of the domain remains unchanged by the transformation such that $\Omega_+=\chi_+$.
The cloak is shaded in dark grey in the deformed configuration.
}
\end{figure}
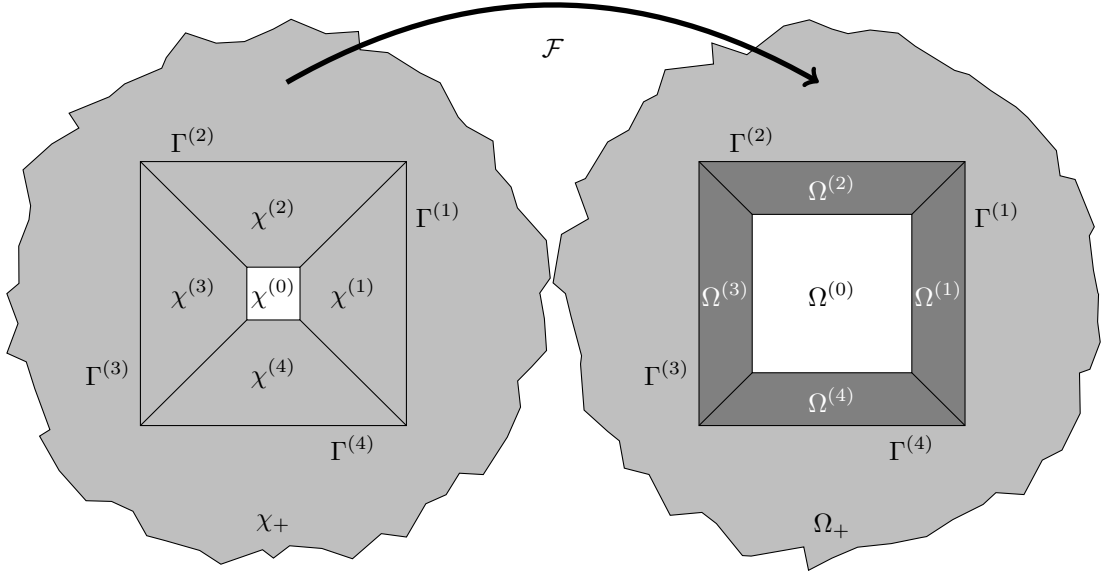

This section is devoted to the construction of a square cloak for flexural waves in a Kirchhoff-Love plate.
Square cloaks have already been constructed for electromagnetic \cite{rahm2008} and out-of-plane elastic \cite{colquitt2013} waves; the cloak presented here is based on the coordinate transformation as in~\cite{colquitt2013}.
Geometrically, the coordinate transformation deforms a small square $\chi^{(0)} = \{\vec{X}: |X_1|/a < \epsilon, |X_2|/a < \epsilon\}$ together with the surrounding four trapezoids $\chi_i$, into a larger square $\Omega^{(0)}=\{\vect{x}: |x_1| < a, |x_2| < a\}$ and four narrower trapezoids $\Omega^{(i)}$ as illustrated in figure~\ref{fig:deformation}. 
The invisibility cloak is then formed from the four deformed trapezoids $\Omega_- = \cup_{i=1}^4 \Omega^{(i)}$ and surrounds the cloaked region $\Omega^{(0)}$.
The map transforms the original domain $\chi = \left(\cup_{i=0}^4 \chi^{(i)}\right)\cup \chi_+$ to the deformed configuration $\Omega = \Omega_+\cup\Omega_-$ in such a way that leaves the exterior of the cloak unchanged, that is, $\mathcal{F}(\chi_+) = \chi_+ = \Omega_+$ and $\mathcal{F}(\Gamma^{(i)}) = \Gamma^{(i)}$.

The mapping is defined in 
such a way that $\mathcal{F}(\vec{X}) = \mathcal{F}^{(i)}(\vec{X})$ for $\vec{X}\in\chi^{(i)}$  ($i=1\ldots4$) and is the identity on $\chi_+ = \Omega_+$.
In particular, 
the mapping, the deformation gradient and the Jacobian for the regions $\Omega^{(1)}$, $\Omega^{(3)}$ are
\[
\mathcal{F}^{(1,3)} = \begin{pmatrix}
\alpha_1X_1 \pm \alpha_2\\
\alpha_1X_2 \pm \alpha_2X_2/X_1
\end{pmatrix},\quad
\vec{F}^{(1,3)} = \begin{pmatrix}
\alpha_1 & 0 \\
\\
\dfrac{x_2\alpha_1\alpha_2}{x_1(\alpha_2\mp x_1)} & \dfrac{x_1 \alpha_1}{x_1\mp\alpha_2}
\end{pmatrix},
\quad
J^{(1,3)} = \frac{x_1\alpha_1^2}{x_1\mp\alpha_2},
\]
where $\alpha_1 = b/(a+b-\epsilon)$, $\alpha_2 = (a+b)(a-\epsilon)/(a+b-\epsilon)$, and $b$ is the thickness of the cloak.
Similarly, for $\Omega^{(2)}$, $\Omega^{(4)}$,
\[
\mathcal{F}^{(2,4)} = \begin{pmatrix}
\alpha_1X_1 \pm \alpha_2X_1/X_2\\
\alpha_1X_2 \pm \alpha_2
\end{pmatrix},\quad
\vec{F}^{(2,4)} = \begin{pmatrix}
\dfrac{x_2 \alpha_1}{x_2\mp\alpha_2} & \dfrac{x_1\alpha_1\alpha_2}{x_2(\alpha_2\mp x_2)}
\\
0 & \alpha_1 \\
\end{pmatrix},
\quad
J^{(2,4)} = \frac{x_2\alpha_1^2}{x_2\mp\alpha_2}.
\]

\subsection{Material parameters and pre-stress for the cloak}
\label{sec:mat-props}

The corresponding material parameters and forces are obtained from the general formalism~\eqref{eq:mat-params}.
For the right hand side of the cloak, the six independent components of flexural rigidity are
\[
\begin{aligned}
D^{(1)}_{1111}  =  \alpha_1^2\left(1-\frac{\alpha_2}{x_1}\right)D^{(0)},\;
D^{(1)}_{2222}  = \frac{\alpha_1^2\left(\alpha_2^2x_2^2+x_1^4\right)^2}{\left(x_1-\alpha_2\right)^3x_1^5}D^{(0)},\;
D^{(1)}_{2211}  = \frac{\alpha_1^2\left(\alpha_2^2x_2^2+x_1^4\right)}{\left(x_1-\alpha_2\right)x_1^3}D^{(0)},\\
D^{(1)}_{1212}  = \frac{\alpha_1^2\alpha_2^2x_2^2}{\left(x_1-\alpha_2\right)x_1^3}D^{(0)},\;
D^{(1)}_{1112}  = -\alpha_1^2\alpha_2\frac{x_2}{x_1^2}D^{(0)},\;
D^{(1)}_{2212}  = -\frac{\alpha_1^2\alpha_2x_2\left(\alpha_2^2x_2^2+x_1^4\right)}{\left(x_1-\alpha_2\right)^2x_1^4}D^{(0)},
\end{aligned}
\]
The remaining components can be deduced from the major and minor symmetries of $\vec{D}$.
The membrane and body forces are
\[
N^{(1)}_{11} = \frac{2\alpha_1^2\alpha_2}{x_1^2\left(x_1-\alpha_2\right)}D^{(0)},\;
N^{(1)}_{12} = \frac{2\alpha_1^2\alpha_2x_2\left(3x_1-2\alpha_2\right)}{\left(x_1-\alpha_2\right)^2x_1^3}D^{(0)},
\]
\[
N^{(1)}_{22} = -\frac{2\alpha_1^2\alpha_2\left(x_1^4+8\alpha_2x_2^2x_1-3\alpha_2^2x_2^2\right)}{x_1^4\left(x_1-\alpha_2\right)^3}D^{(0)},
\]
\[
S^{(1)}_1 = 0,\quad S^{(1)}_2 = \frac{24\alpha_1^2\alpha_2 x_2}{\left(x_1-\alpha_2\right)^3x_1^2}D^{(0)}.
\]
Finally, the density is
\[
\rho^{(1)} = \frac{\Rho\left(x_1-\alpha_2\right)}{\alpha_1^2x_1}.
\]
The corresponding physical quantities for the remaining three sides of the cloak are provided in appendix~\ref{ap:mat-params}.
The material on the interior of the cloak corresponds to an inhomogeneous anisotropic Kirchhoff-Love plate with inhomogeneous and anisotropic pre-stress.
All material parameters and forces are finite for $\epsilon>0$.

\subsubsection{Principal directions of orthotropy}

\begin{figure}
\centering
\includegraphics[width=0.45\linewidth]{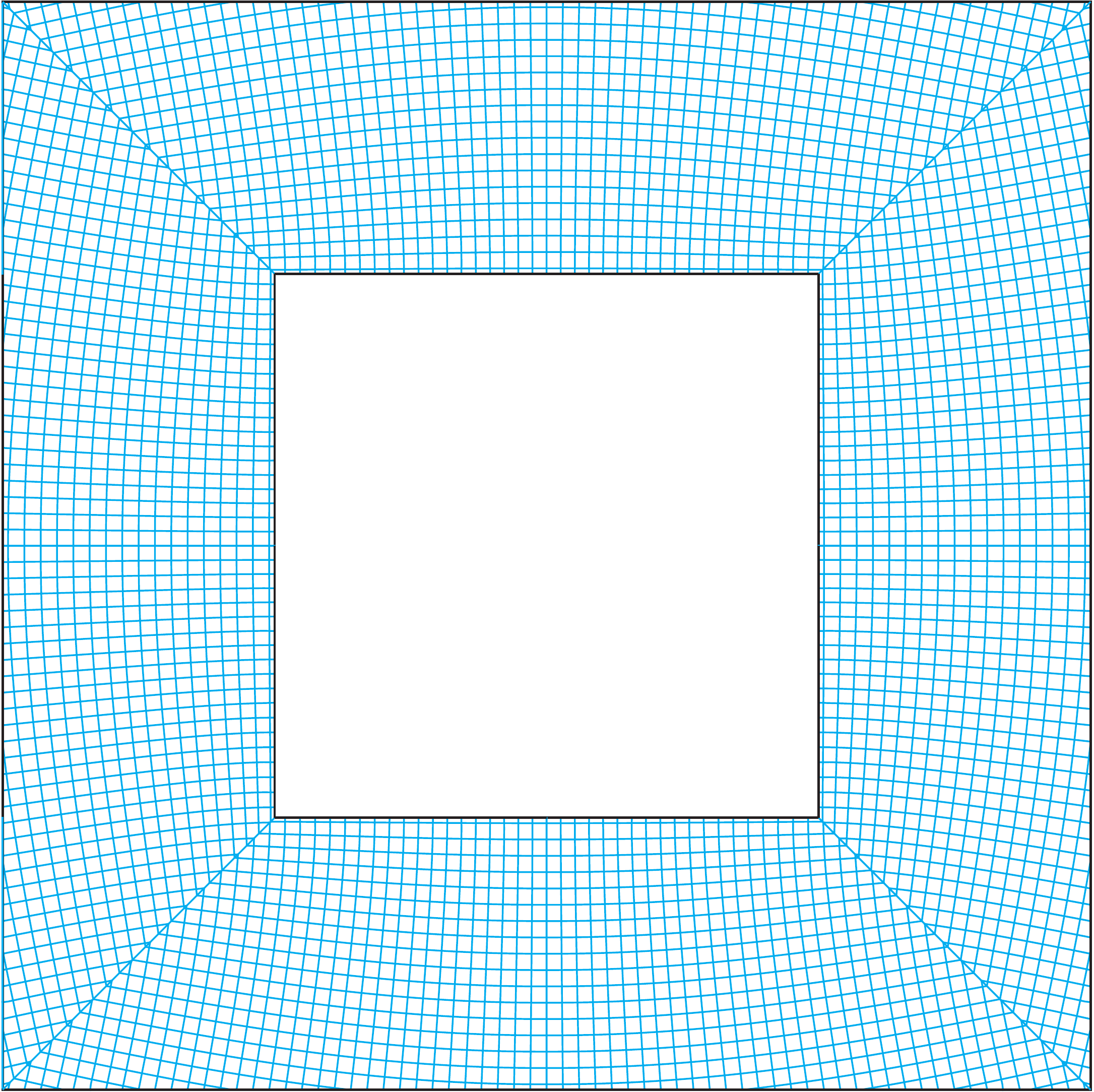}
\caption{\label{fig:grid-plate}
The principal directions of the flexural rigidity tensor for the square cloak in a Kirchhoff-Love plate.
Results are presented for parameter values: $a=0.5$ m, $b=0.5$ m and $\epsilon=10^{-3}$.
}
\end{figure}

At each point of the plate, one can introduce a system of coordinates, which coincides with the principal directions of orthotropy of the plate.
To this end, a local angle of rotation $\theta$ is introduced. 
The rotation is considered local in the sense that the angle of rotation $\theta=\theta(\vect{x})$ is a function of position.
For convenience, the following reduced form for the six independent components of flexural rigidities are introduced
\[
\begin{aligned}
D^{(i)}_{11} & = D^{(i)}_{1111},\;
&D^{(i)}_{12} & = D^{(i)}_{1122},\;
&D^{(i)}_{16} & = D^{(i)}_{1112},\\
D^{(i)}_{22} & = D^{(i)}_{2222},\;
&D^{(i)}_{26} & = D^{(i)}_{2221},\;
&D^{(i)}_{66} & = D^{(i)}_{1212}.
\end{aligned}
\]
The reduced flexural rigidities $D^{(i)}_{16}$ and $D^{(i)}_{26}$, sometimes called \emph{auxiliary rigidities}, vanish in the principal directions of the plate (see~\cite{timoshenko1959, lekhnitskii1968} among others).
Consider a \emph{local} rotation of the coordinate system $\vect{\mathcal{G}}: \vect{x}\mapsto\tilde{\vect{x}}$.
In this new coordinate system, the auxiliary rigidities may be expressed as
\begin{equation}
\begin{aligned}
\tilde{D}^{(i)}_{16} & = \frac{1}{2}\left[D^{(i)}_{22}\sin^2\theta - D^{(i)}_{11}\cos^2\theta + \left(D^{(i)}_{12} + 2D^{(i)}_{66}\right)\cos2\theta\right]\sin2\theta \\
 & \qquad\qquad
 +D^{(i)}_{16}\cos^2\theta\left(\cos^2\theta-3\sin^2\theta\right) + D^{(i)}_{26}\sin^2\theta\left(3\cos^2\theta-\sin^2\theta\right),\\
 \tilde{D}^{(i)}_{26} & = \frac{1}{2}\left[D^{(i)}_{22}\cos^2\theta - D^{(i)}_{11}\sin^2\theta - \left(D^{(i)}_{12} + 2D^{(i)}_{66}\right)\cos2\theta\right]\sin2\theta \\
 & \qquad\qquad
 +D^{(i)}_{16}\sin^2\theta\left(3\cos^2\theta-\sin^2\theta\right) + D^{(i)}_{26}\cos^2\theta\left(\cos^2\theta-3\sin^2\theta\right),
\end{aligned}
\label{eq:rot-D}
\end{equation}
where the dependence of $\theta$ and $D^{(i)}_{jk}$ on $\vect{x}$ has been omitted but is understood.
The appropriate local angles of rotation $\theta(\vect{x})$, which satisfy the above system of transcendental equations yields the local principal directions of rigidity.
These principal directions for the square push-out transformation applied to the equations governing the flexural displacement of a Kirchhoff-Love plate are shown in figure~\ref{fig:grid-plate}.

If $D^{(i)}_{16}=D^{(i)}_{26}=0$, then the system of coordinates is already aligned with principal directions of orthotropy $(\theta=0)$. If additional non-zero rotation is needed $(\theta\neq 0)$, then equations \eqref{eq:rot-D} can be written in the form :
\begin{equation}
\begin{aligned}
\tilde{D}^{(i)}_{16}+\tilde{D}^{(i)}_{26} & = \frac{1}{2}\left(D^{(i)}_{22}- D^{(i)}_{11}\right)\sin2\theta +
\left(D^{(i)}_{16}+D^{(i)}_{26}\right)\cos2\theta,\\
\tilde{D}^{(i)}_{26}-\tilde{D}^{(i)}_{16} & = \frac{1}{2}\left(D^{(i)}_{11}+D^{(i)}_{22} - 2D^{(i)}_{12} - 4D^{(i)}_{66}\right)\sin2\theta\cos2\theta 
 +\left(D^{(i)}_{26}- D^{(i)}_{16}\right)\left(\cos^22\theta-\sin^22\theta\right),
\end{aligned}
\label{eq:rot-D sum diff}
\end{equation}
which have to be set equal to zero in order to find the principal direction angle $\theta\in(0,\pi/2]$.
Substituting the solution of the first equation in (\ref{eq:rot-D sum diff}) 
\begin{equation}
\sin2\theta = 2\frac{D^{(i)}_{16}+D^{(i)}_{26}}{D^{(i)}_{11}- D^{(i)}_{22}}\cos 2\theta,
\label{eq:rot-D cond1}
\end{equation}
into the second one, leads to
\begin{equation}
\cos4\theta =-\frac{\left(D^{(i)}_{11}+D^{(i)}_{22} - 2D^{(i)}_{12} - 4D^{(i)}_{66}\right)\left(D^{(i)}_{16}+D^{(i)}_{26}\right)}
{\left(D^{(i)}_{11}+D^{(i)}_{22} - 2D^{(i)}_{12} - 4D^{(i)}_{66}\right)\left(D^{(i)}_{16}+D^{(i)}_{26}\right)+2
\left(D^{(i)}_{11}- D^{(i)}_{22}\right)\left(D^{(i)}_{26}-D^{(i)}_{16}\right)}.
\label{eq:rot-D sol1}
\end{equation}
It is also observed that, for this angle of rotation, the rigidity component $\tilde{D}^{(i)}_{66}$ vanishes.

This local orthotropy gives further important physical meaning to the material parameters of the cloak.
In particular, the sign of some flexural rigidities changes across the symmetry lines in the global coordinate system.
This change in sign can now be interpreted as simply a change in orientation of the principal axes of the material. 

It is interesting to compare the principal directions in the plate and in the membrane cloaking problem, considered in~\cite{colquitt2013}, where the same geometric transformation is used.
As shown in~\cite{colquitt2013}, the principal directions of orthotropy for a membrane cloak are characterised by the eigenvectors of $\vec{G}$ (cf. equation~\eqref{eq:Gij}).
If we identify the local angle between the eigenvectors of $\vec{G}$ and the standard Cartesian basis vectors with $\theta$, it is straightforward to verify that such an angle satisfies eqns.~\eqref{eq:rot-D}, and thus, the principal directions of orthotropy for the membrane cloaking problem are the same as those for the Kirchhoff-Love cloaking problem. 

\subsection{Implementation of the cloak for the flexural plate}
\label{sec:sims}

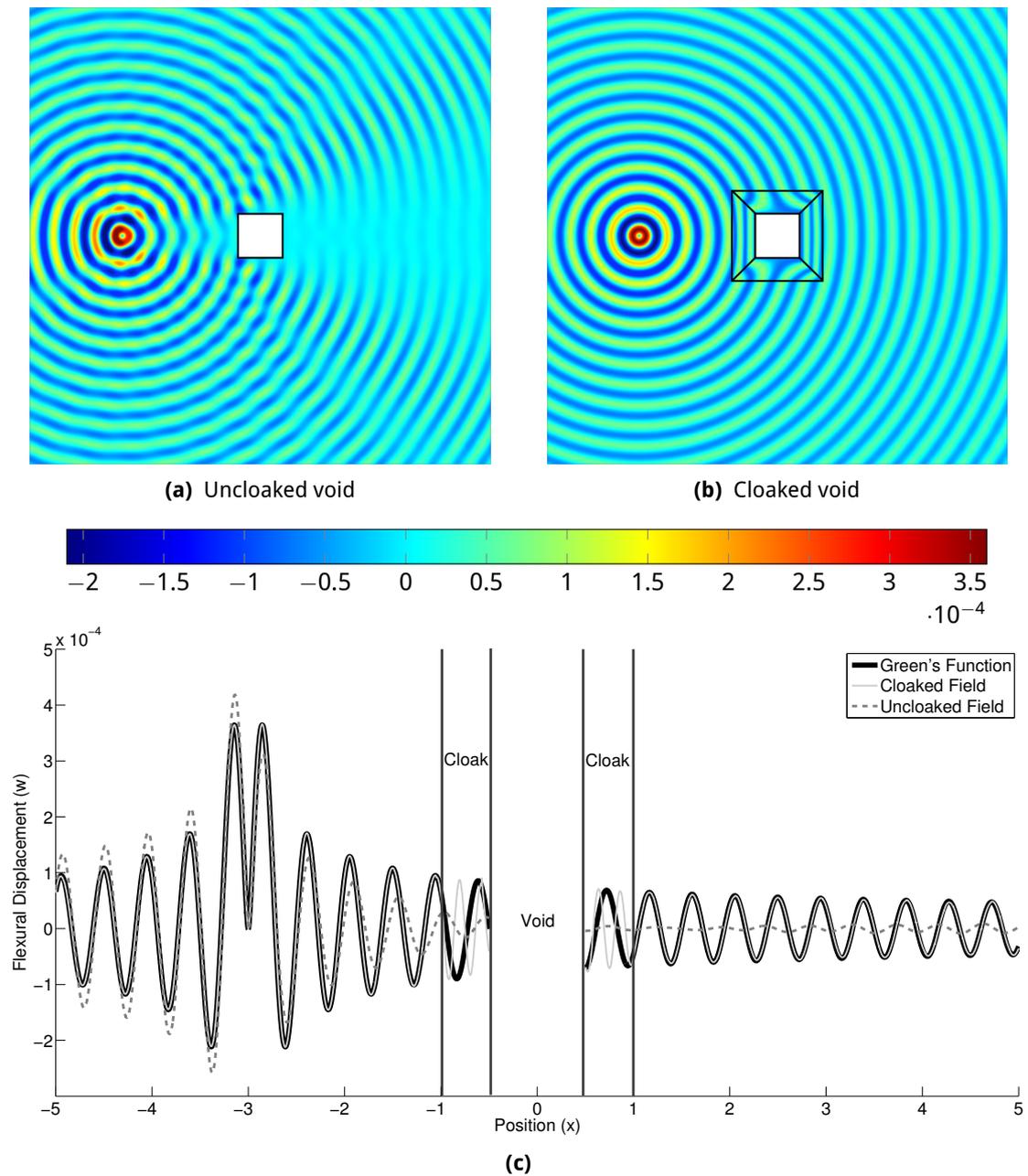
\begin{figure}
\centering
\begin{subfigure}[c]{0.45\textwidth}
\includegraphics[width=\linewidth]{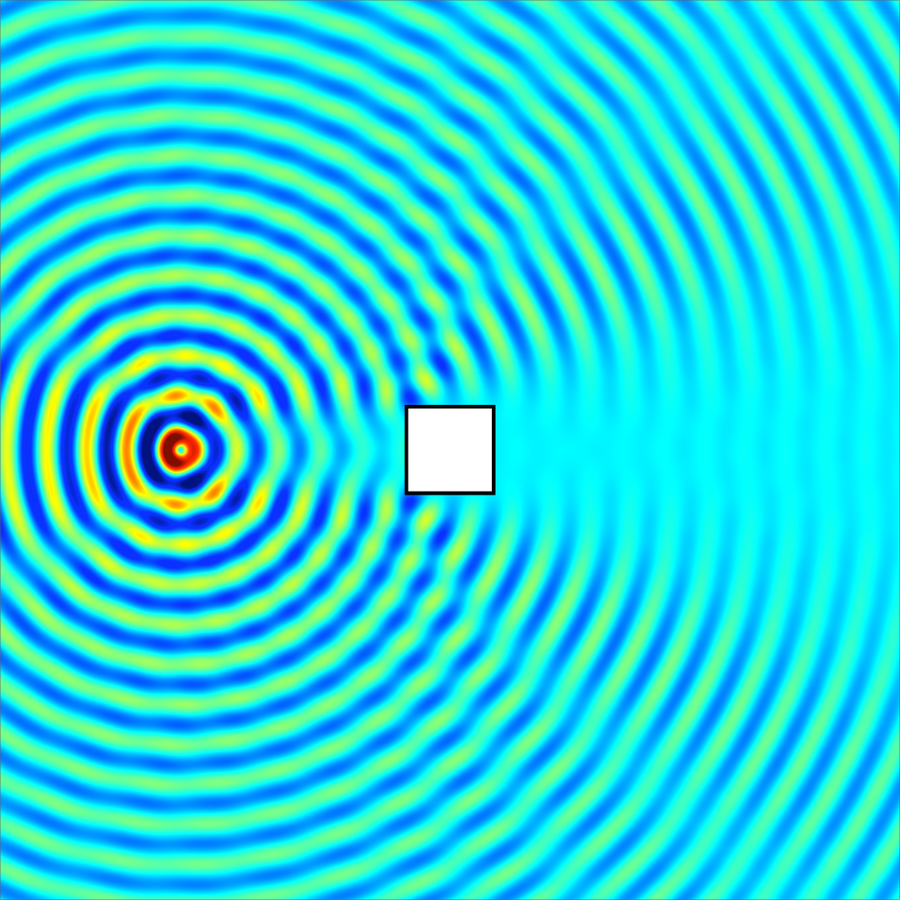}
\caption{\label{fig:Uncloaked}
Uncloaked void}
\end{subfigure}
\qquad
\begin{subfigure}[c]{0.45\textwidth}
\includegraphics[width=\linewidth]{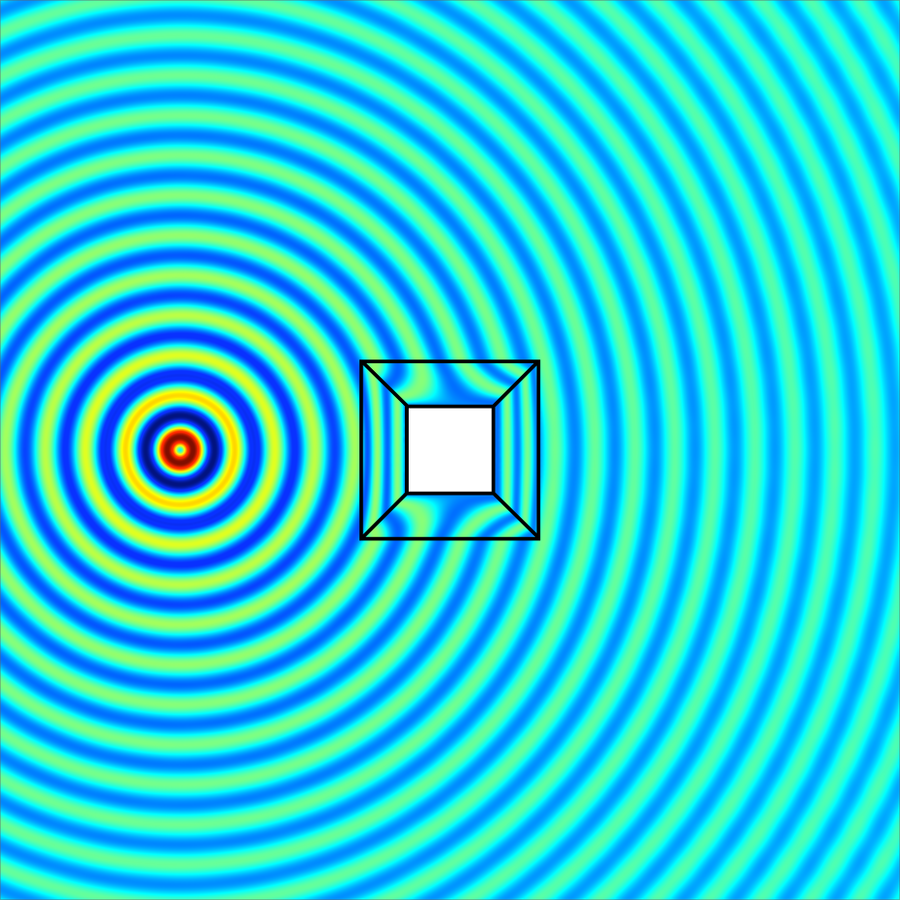}
\caption{\label{fig:Cloaked}
Cloaked void}
\end{subfigure}
\begin{subfigure}[c]{0.9\textwidth}
\centering
\begin{tikzpicture}
\begin{axis}[
    hide axis,
    scale only axis,
    height=0pt,
    width=0pt,
    colormap/jet,
    colorbar horizontal,
    point meta min=-0.00021,
    point meta max=0.00036,
    colorbar style={
        width=\linewidth,
    }]
    \addplot [draw=none] coordinates {(0,0)};
\end{axis}
\end{tikzpicture}
\end{subfigure}
\qquad
\begin{subfigure}[c]{\textwidth}
\includegraphics[width=\linewidth]{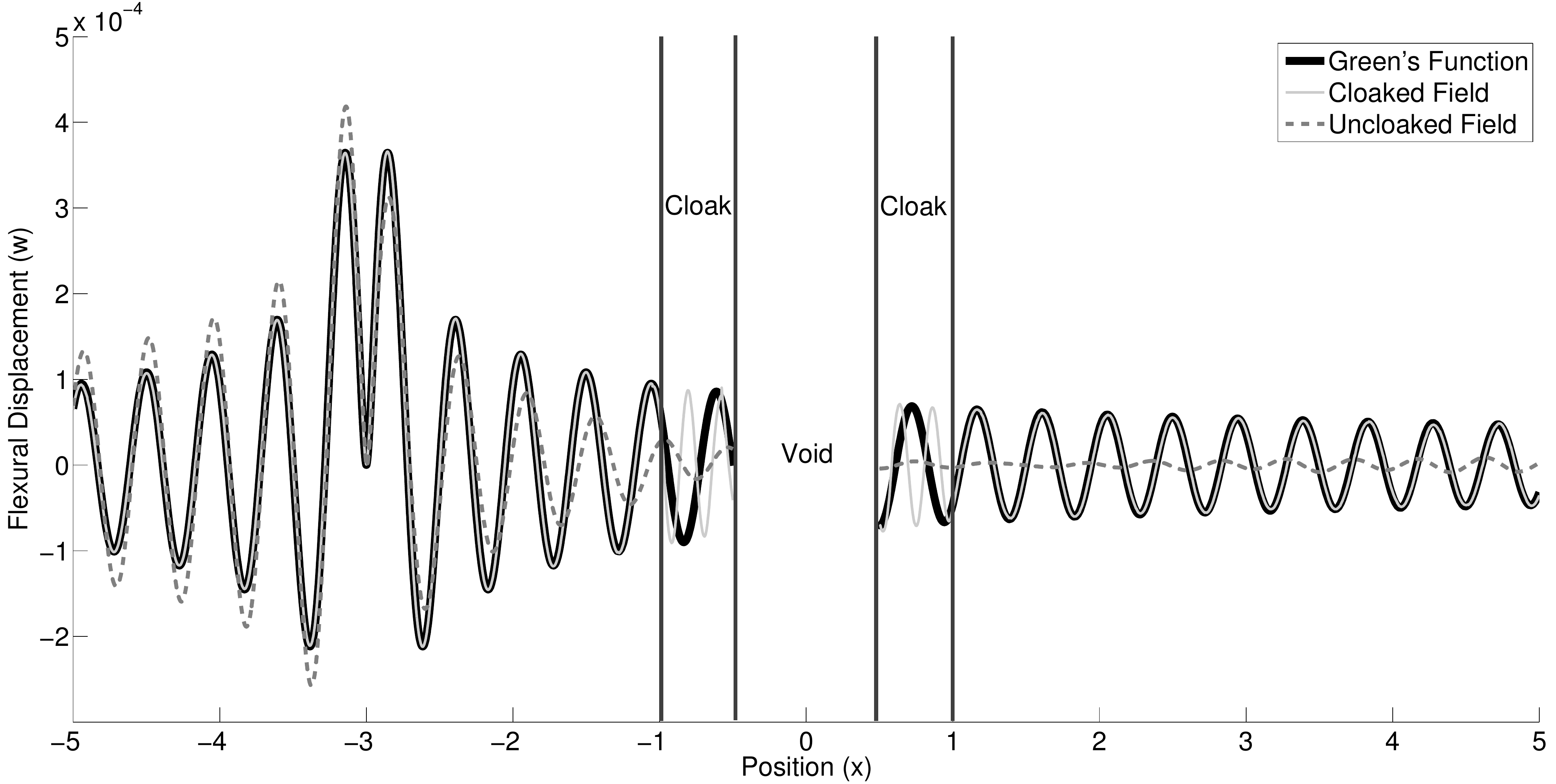}
\caption{\label{fig:Cloaked-Uncloaked-Line}}
\end{subfigure}
\caption{\label{fig:Cloaking}
The flexural displacement $w(\vect{x})$ generated by a point source in the presence of an uncloaked (a) and cloaked (b) void.
Figure (c) shows the flexural displacement for cases (a) \& (b) together with the Green's function for a Kirchhoff-Love plate along the line passing through the source and centre of the cloak.
The radian frequency is $\omega=314$ rad/s.}
\end{figure}

In order to simulate a cloak around a scatterer  in an  infinite Kirchhoff-Love plate, perfectly matched layers are used in the vicinity of the exterior boundary of the computational domain. The implementation uses a bespoke algorithm  developed for the push-out cloaking transformation in  the finite element software COMSOL Multiphysics\textsuperscript{\textregistered}. 
A steel plate of thickness $10^{-3}$ m is considered and the following parameter values are chosen: $D^{(0)} =19.23$ Nm, $ \Rho =7800$ kg/m$^3$,  $h = 10^{-3}$ m, $a=b=0.5$ m, $\epsilon=10^{-3}$ and $\omega = 314$ rad/s. 
It is emphasised that these parameters are used to demonstrate broadband cloaking, which persists over a wide range of frequencies, geometrical and material parameter values, within the usual constraints of the Kirchhoff-Love plate model.
Figure~\ref{fig:Uncloaked} shows 
the flexural displacement generated by a cylindrical source in a Kirchhoff-Love plate with a square void.
Figure~\ref{fig:Cloaked} shows the corresponding field when the void is surrounded by a square cloak, constructed as described in Section~\ref{sec:example-cloak}.
Figure~\ref{fig:Cloaked-Uncloaked-Line} shows the fields for plots~\ref{fig:Uncloaked} and~\ref{fig:Cloaked} together with Green's function for an infinite Kirchhoff-Love plate along the line passing through the source and centre of the cloak.
Here, the Green's function represents the unperturbed field in the absence of the void and cloak.
it is observed that, outside the cloaked region, the cloaked field and unperturbed field are almost coincident.
Figure~\ref{fig:Cloaking} clearly illustrates the effectiveness of this platonic cloak.

\subsubsection{Green's functions and comparison with cloaking for the Helmholtz operator}
\label{sec:corresponding-helmholtz}

It is clear from the transformed equation~\eqref{eq:transformed-plate-eq-index} that, in general, cloaks for Kirchhoff-Love plates cannot be decomposed into Helmholtz and modified-Helmholtz parts.
Physically this means that, within the cloak, propagating and evanescent modes couple.
However, if the ambient plate on the exterior of the cloak is homogeneous one would expect propagating modes, corresponding to solutions of the Helmholtz equation, to dominate outside the immediate neighbourhood of the cloak.
The corresponding problem for the membrane (Helmholtz operator) and a square cloak was recently considered by Colquitt et al.~\cite{colquitt2013} and in the context of electromagnetism by Rahm et al.~\cite{rahm2008}.
In Cartesian coordinates the transformed Helmholtz equation, which governs the out-of-plane displacement of a membrane, may be expressed as
\[
G_{ij}u_{,ij} + G_{ij,i} u_{,j} + \frac{\Rho_m\omega^2}{J\mu^{(0)}} u = 0,
\]
where $\mu^{(0)}$ is the stiffness of the untransformed membrane and $\Rho_m$ the density.
It is immediately apparent that, in general, solutions of the transformed Helmholtz equation will not satisfy the transformed equation for a plate~\eqref{eq:transformed-plate-eq-index}.
Nevertheless, if the exterior medium is homogeneous, one would expect solutions of the ``corresponding'' membrane problem to dominate in the plate solution at infinity.
The Green's function for equation~\eqref{eq:untransformed-plate-eq}, which governs the flexural displacement of a Kirchhoff-Love plate, is~\cite{evans2007}
\begin{equation}
g_P(\vect{x}) = i\frac{\Hankel_0^{(1)}(i\beta|\vect{x}|)-\Hankel_0^{(1)}(\beta|\vect{x}|)}{8\beta^2D^{(0)}},
\label{eq:GF-plate}
\end{equation}
where $\beta^4 = \omega^2\Rho h/D^{(0)}$.
The Green's function for the two-dimensional Helmholtz equation, which governs the out-of-plane displacement of a membrane, is
\begin{equation}
g_H(\vect{x}) = -i\frac{\Hankel_0^{(1)}(k|\vect{x}|)}{4\mu^{(0)}},
\label{eq:GF-helmholtz}
\end{equation}
where $k^2 = \omega^2\Rho_m/\mu^{(0)}$.
For large arguments, the Hankel function $\Hankel_0^{(1)}$ has the following asymptotic representation~\cite{olver2010}
\[
\Hankel_0^{(1)}(z) \sim \sqrt{\frac{2}{\pi z}}e^{i\left(z-\frac{\pi}{4}\right)},
\]
as $z\to\infty$ in $-\pi + \delta \leq \arg(z) \leq 2\pi-\delta$, where $0<\delta\ll 1$.
Using the above representation yields the following expressions for the Green's functions at infinity,
\[
g_P(\vect{x}) \sim -\frac{i}{8\beta^2 D^{(0)}}\sqrt{\frac{2}{\pi\beta|\vect{x}|}}e^{i\left(\beta|\vect{x}| - \frac{\pi}{4}\right)}
\quad\text{and}\quad
g_H(\vect{x}) \sim -\frac{i}{4\mu^{(0)}}\sqrt{\frac{2}{\pi k|\vect{x}|}}e^{i\left(k|\vect{x}| - \frac{\pi}{4}\right)},
\]
whence is clear that $\beta=k$ is required for the two fields to share the same phase in the far field and choosing $2\beta^2D^{(0)} = \mu^{(0)}$ gives equal amplitudes.
Thus, for the cloaking problem for the plate and Helmholtz equation to have the same solution in the far field, 
the material parameters should be chosen such that $\Rho_m= 2\Rho h$ and $\mu^{(0)} = 2\omega\sqrt{\Rho h D^{(0)}}$.
It is in this sense that the cloaking problem for the membrane is said to ``correspond'' to the cloaking problem for the plate.

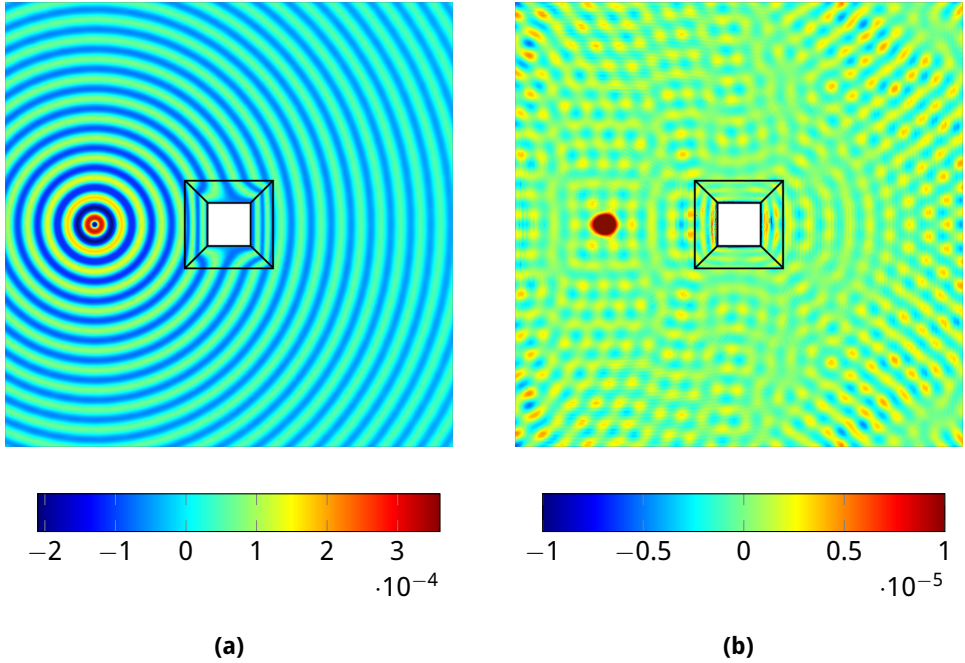
\begin{figure}
\centering
\begin{subfigure}[c]{0.4\textwidth}
\includegraphics[width=\linewidth]{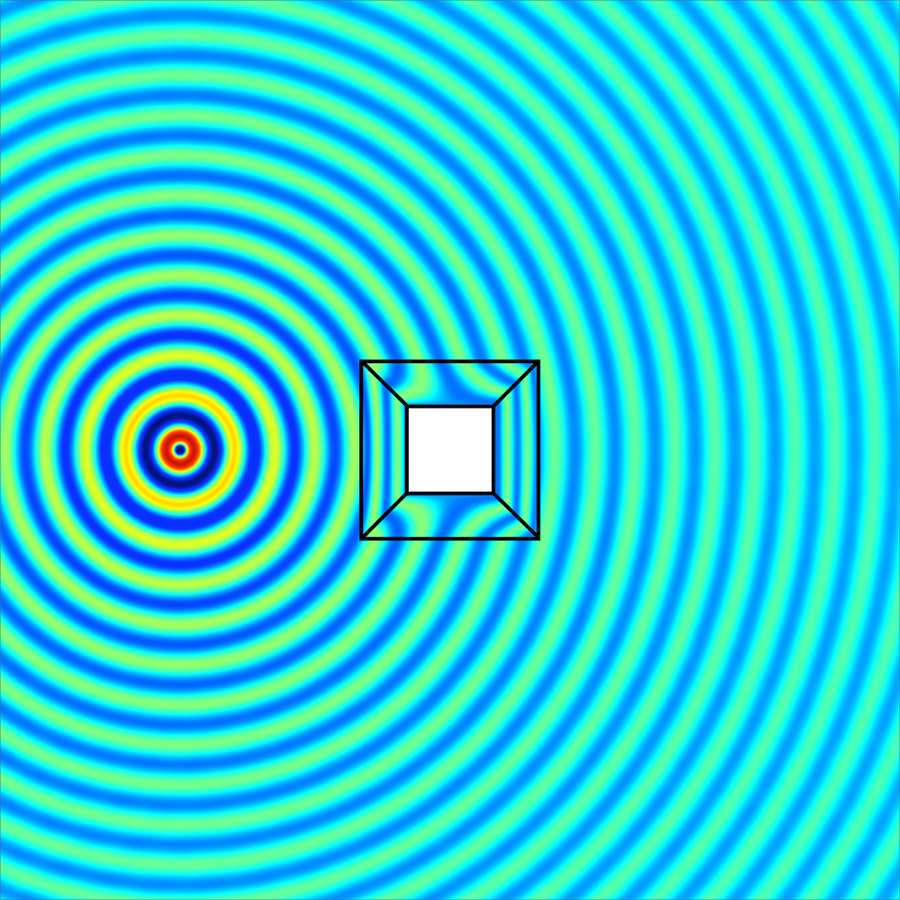}
\begin{center}
\begin{tikzpicture}
\begin{axis}[
    hide axis,
    scale only axis,
    height=0pt,
    width=0pt,
    colormap/jet,
    colorbar horizontal,
    point meta min=-0.00021,
    point meta max=0.00036,
    colorbar style={
        width=0.9\linewidth,
    }]
    \addplot [draw=none] coordinates {(0,0)};
\end{axis}
\end{tikzpicture}
\end{center}
\caption{\label{fig:Helmholtz-Real}}
\end{subfigure}
\qquad
\begin{subfigure}[c]{0.4\textwidth}
\includegraphics[width=\linewidth]{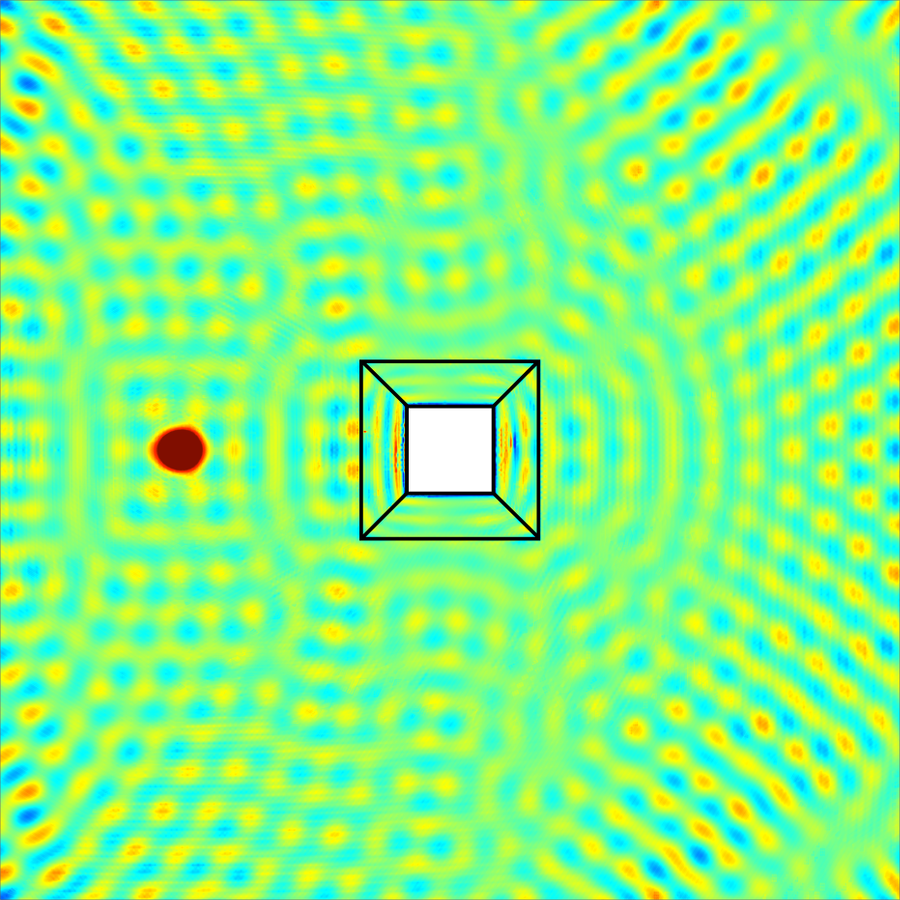}
\begin{center}
\begin{tikzpicture}
\begin{axis}[
    hide axis,
    scale only axis,
    height=0pt,
    width=0pt,
    colormap/jet,
    colorbar horizontal,
    point meta min=-0.00001,
    point meta max=0.00001,
    colorbar style={
        width=0.9\linewidth,
    }]
    \addplot [draw=none] coordinates {(0,0)};
\end{axis}
\end{tikzpicture}
\end{center}
\caption{\label{fig:Difference-Real}}
\end{subfigure}
\caption{\label{fig:Difference-Helmholtz-Plate}
(a) The solution for the corresponding membrane cloaking problem, (b) The difference between the solution for the cloaking problem for plates and the corresponding membrane problem.}
\end{figure}

\begin{figure}
\centering
\begin{subfigure}[c]{0.8\textwidth}
\includegraphics[width=\linewidth]{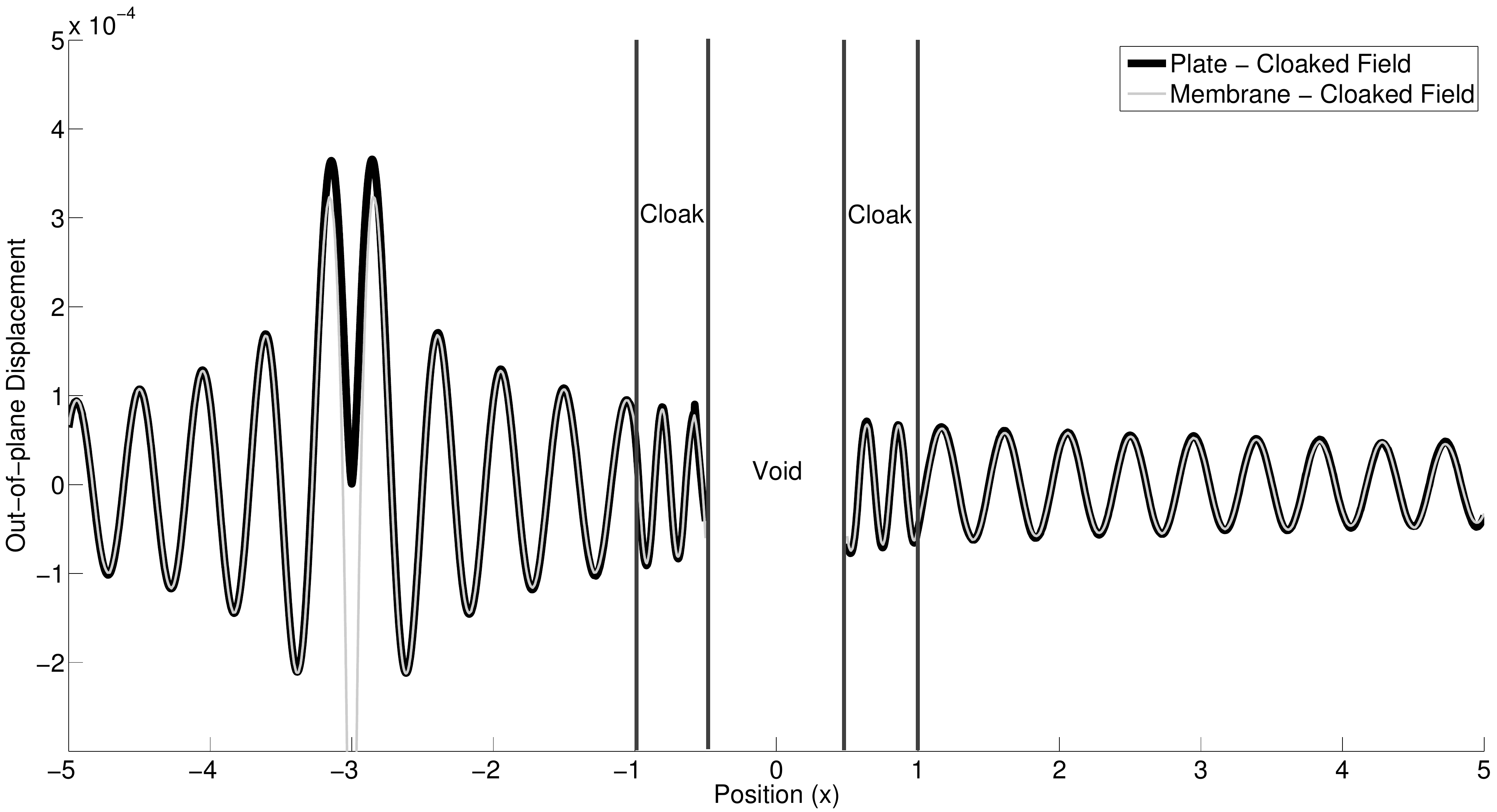}
\caption{\label{fig:Line-Comp-Helmholtz-Plate}}
\end{subfigure}
\begin{subfigure}[c]{0.8\textwidth}
\includegraphics[width=\linewidth]{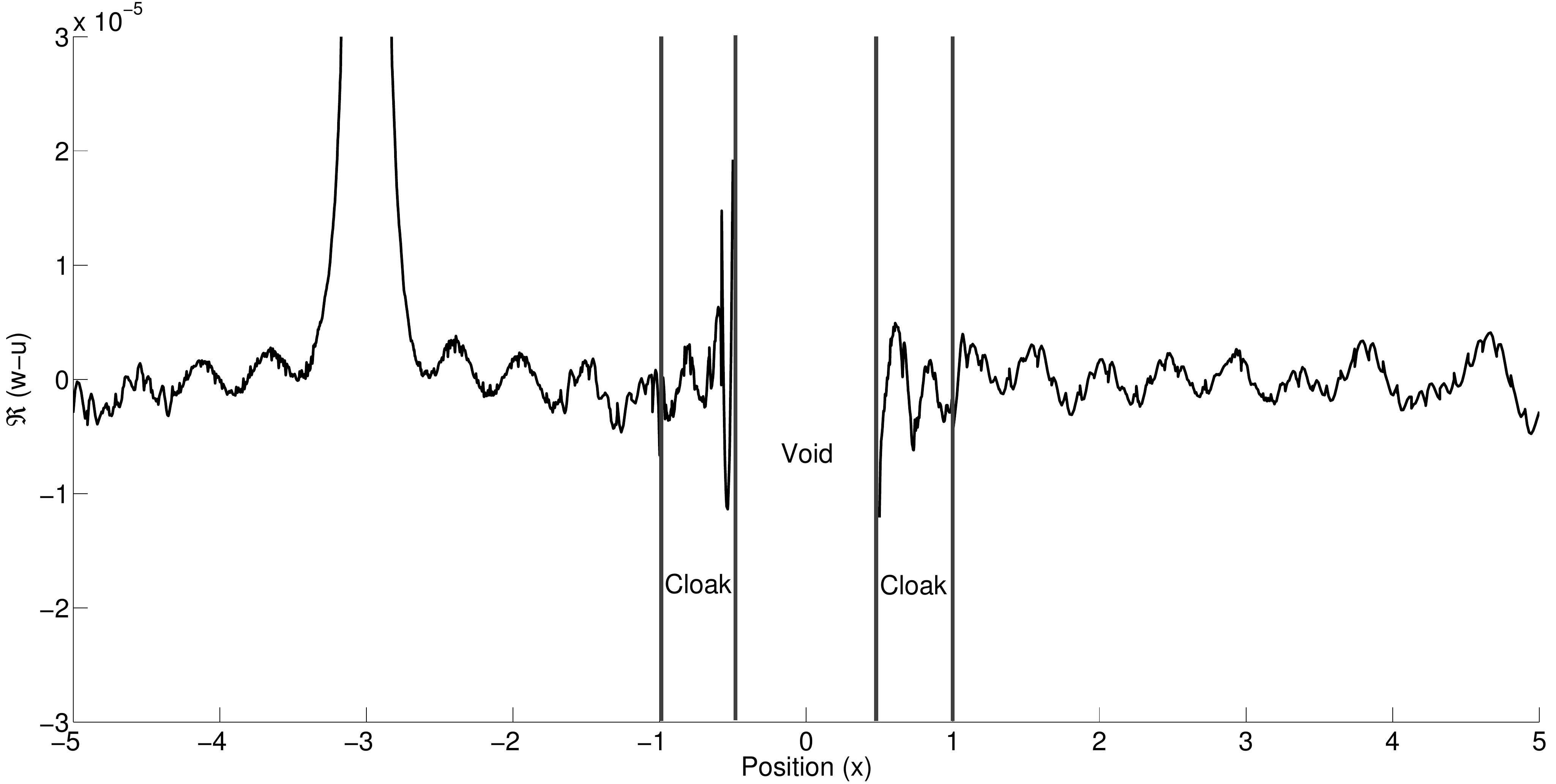}
\caption{\label{fig:Line-Diff-Helmholtz-Plate}}
\end{subfigure}
\caption{\label{fig:Difference-Helmholtz-Plate-Line}
Figure (a) shows the solution when considering the cloaking problem for plates and the corresponding membrane problem.
Figure (b) shows the difference between the solutions when considering the cloaking problem for plates and the corresponding membrane problem.
In figures (a) \& (b), the fields are plotted along the line passing through the source and centre of the cloak.
The reader's attention is drawn to the different scales in figures (a) and (b).}
\end{figure}

Figure \ref{fig:Difference-Helmholtz-Plate} shows the solution of the cloaking problem for the corresponding membrane problem (see~\cite{colquitt2013}), and the difference between the fields for the plate and membrane.
In figure~\ref{fig:Line-Comp-Helmholtz-Plate}, the two solutions for the two cloaking problems (plate and membrane) are plotted along the line passing through the source and centre of the cloak.
Figure \ref{fig:Line-Diff-Helmholtz-Plate} shows the difference between the two solutions along the same line.
The reader's attention is drawn to the different scales in figures \ref{fig:Cloaking}, \ref{fig:Difference-Helmholtz-Plate}  and \ref{fig:Difference-Helmholtz-Plate-Line}.
It is also emphasised that the Green's function for the membrane problem (Helmholtz operator) \eqref{eq:GF-helmholtz} is singular at the origin whereas the fundamental solution for the plate equation \eqref{eq:GF-plate} is regular, hence the large discrepancy in the vicinity of the source in figures \ref{fig:Cloaking},  \ref{fig:Difference-Helmholtz-Plate} and \ref{fig:Difference-Helmholtz-Plate-Line} is justified.
It is observed that, away from the source and outside the cloak, the difference between the solutions for the plate problem and corresponding problem for the membrane is small.
However, this difference is significant on the interior of the cloak, particularly close to the inner boundary of the cloak.

\subsubsection{The quality of cloaking}
\label{sec:quality}

In order to better asses the efficacy of invisibility cloaks, it is desirable to have a quantifiable measure of the \emph{quality} of the cloaking effect.
As remarked in~\cite{colquitt2013}, it is not obvious what \emph{quality} means with respect to a cloak.
Some authors use a scattering cross-section to assess the cloaking effect (see, for example,~\cite{norris2012}).
Experimental studies such as~\cite{stenger2012} have used an $L_2$ norm computed directly from the measured fields in order to quantify the efficiency of the cloak.
For the present work, the scattering measure introduced in~\cite{colquitt2013} will be used as a tool to quantify the effectiveness of the cloak:
\begin{equation}
\mathcal{E} (u_1, u_0,\mathcal{R})
=\left(\int\limits_{\mathcal{R}} |u_1(\vect{x}) - u_0(\vect{x})|^2\;\mathrm{d} \vect{x}\right)
\left(\int\limits_{\mathcal{R}} |u_0(\vect{x})|^2\;\mathrm{d} \vect{x}\right)^{-1},
\end{equation}
where $\mathcal{R}\subset\mathbb{R}^2$ is some region outside the cloak, $u_1(\vect{x})=\Re w(\vect{x})$ is the real part of the measured field, and $u_0(\vect{x}) = \Re g_P(\vect{x})$ is the real part of Green's function for the unperturbed problem and represents the ideal field. Thus, perfect cloaking corresponds to a vanishing $\mathcal{E}$.
For the illustrative simulations shown in figure~\ref{fig:Cloaking}, the region $\mathcal{R}$
is chosen to be the entire computational domain, excluding the cloak and a small disc enclosing the point source.
It is emphasised that this choice of $\mathcal{R}$ yields an extremely strict measure of the efficacy of the cloak; with not only forward and backward scattering effects accounted for, but also significant near field effects in the immediate neighbourhood of the cloak.
The scattering measure for the uncloaked field shown in figure~\ref{fig:Uncloaked} is $\mathcal{E}(w_u, u_0,\mathcal{R}) = 0.109$, whereas the scattering measure for the cloaked field shown in figure~\ref{fig:Cloaked} is $\mathcal{E}(w_c, u_0,\mathcal{R}) = 5.05\times10^{-4}$.
The difference between the two scattering measures serves to emphasise the efficiency of the cloak.

\subsubsection{Measuring cloaking quality for interference patterns}
\label{sec:interference}

Recently, interferometry has been proposed as a possible method through which the quality of cloaks may be assessed.
In particular, Colquitt et al.~\cite{colquitt2013} illustrated the efficacy of a square cloak for Helmholtz waves using a Young's double-slit interferometer.
In~\cite{colquitt2013} the stability of the interference pattern was examined as a cloaked and uncloaked object were successively placed over one of the apertures.
It was demonstrated that there is virtually no perturbation to the interference fringes when an aperture is covered with a cloaked object, whereas the uncloaked object severely distorts the interference pattern.

\begin{figure}
\centering
\begin{subfigure}[c]{0.3\textwidth}
\includegraphics[width=\linewidth]{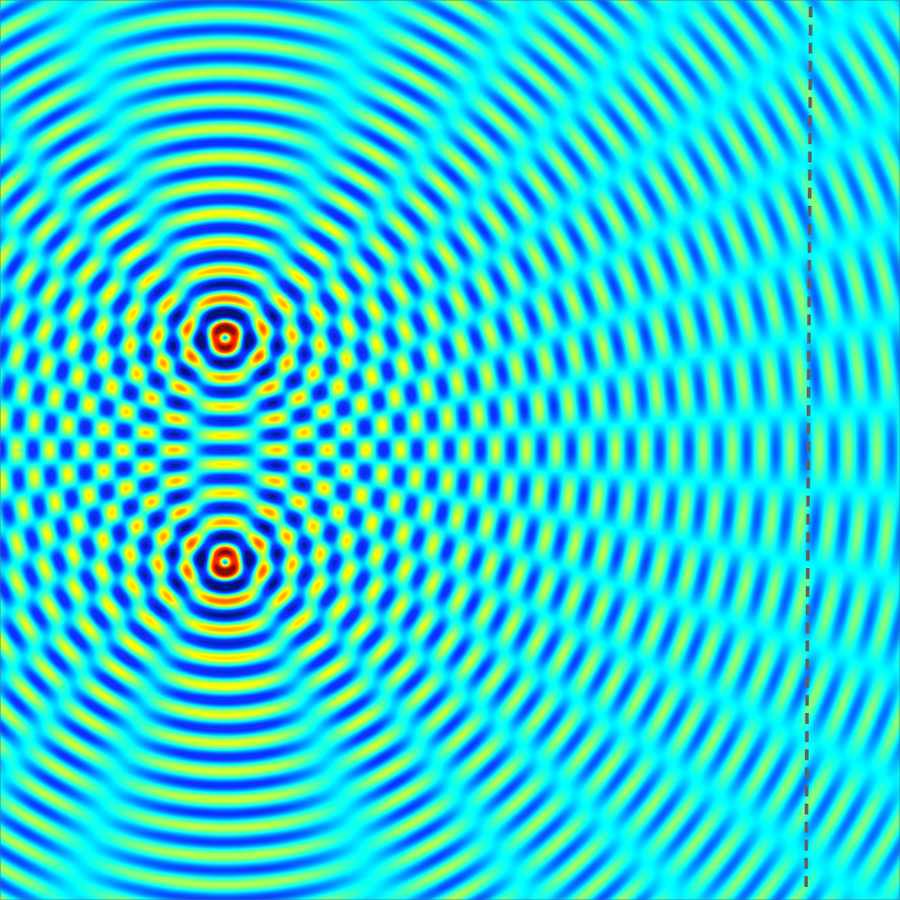}
\caption{\label{fig:Interferometry-Intact}}
\end{subfigure}
\begin{subfigure}[c]{0.3\textwidth}
\includegraphics[width=\linewidth]{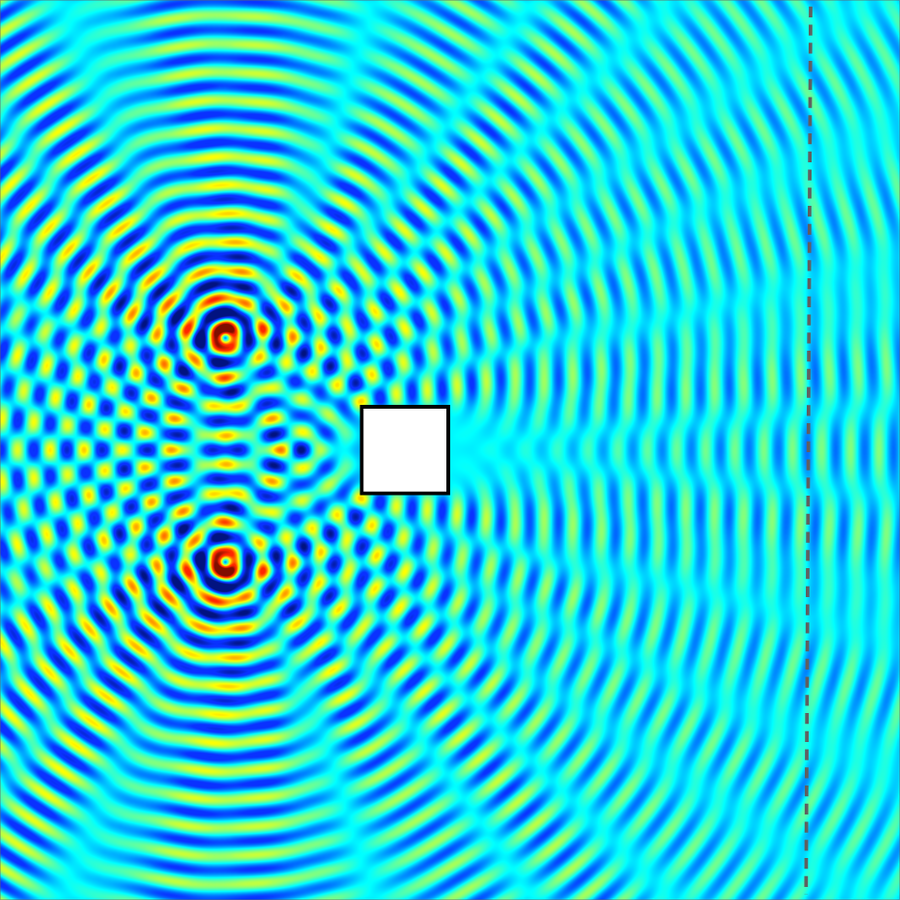}
\caption{\label{fig:Interferometry-Uncloaked}}
\end{subfigure}
\begin{subfigure}[c]{0.3\textwidth}
\includegraphics[width=\linewidth]{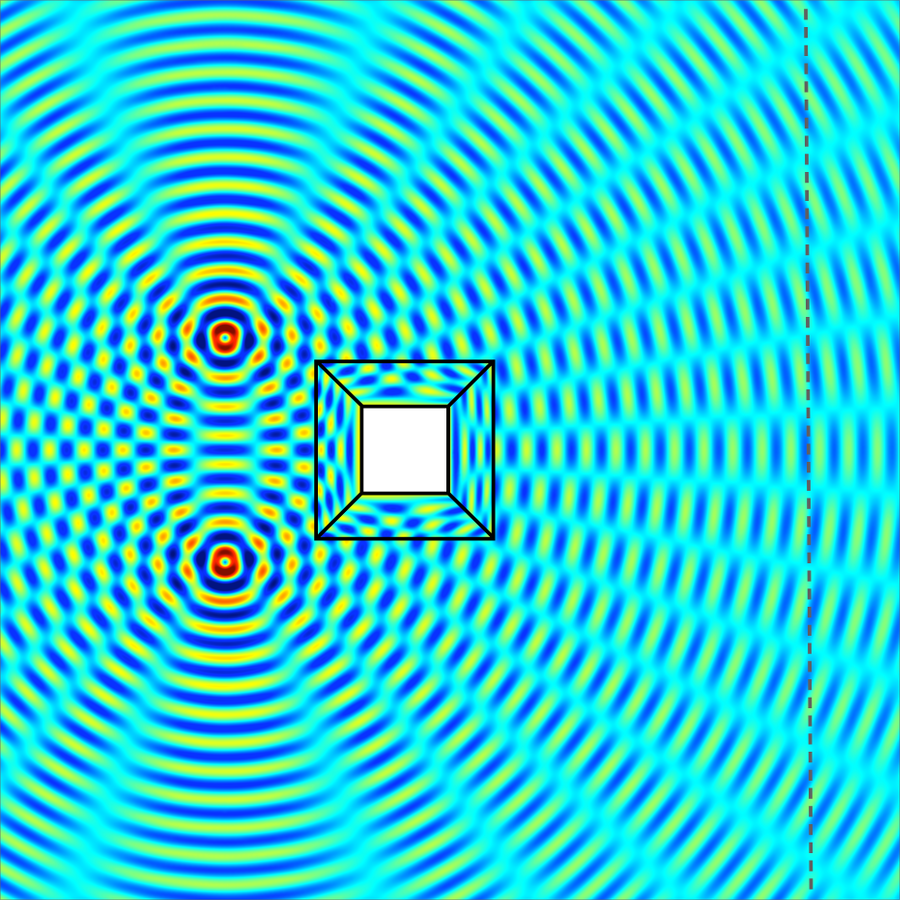}
\caption{\label{fig:Interferometry-Cloaked}}
\end{subfigure}
\begin{subfigure}[c]{\textwidth}
\centering
\begin{tikzpicture}
\begin{axis}[
    hide axis,
    scale only axis,
    height=0pt,
    width=0pt,
    colormap/jet,
    colorbar horizontal,
    point meta min=-0.00042,
    point meta max=0.00072,
    colorbar style={
        width=0.8\linewidth,
    }]
    \addplot [draw=none] coordinates {(0,0)};
\end{axis}
\end{tikzpicture}
\end{subfigure}
\begin{subfigure}[c]{0.9\textwidth}
\includegraphics[width=\linewidth]{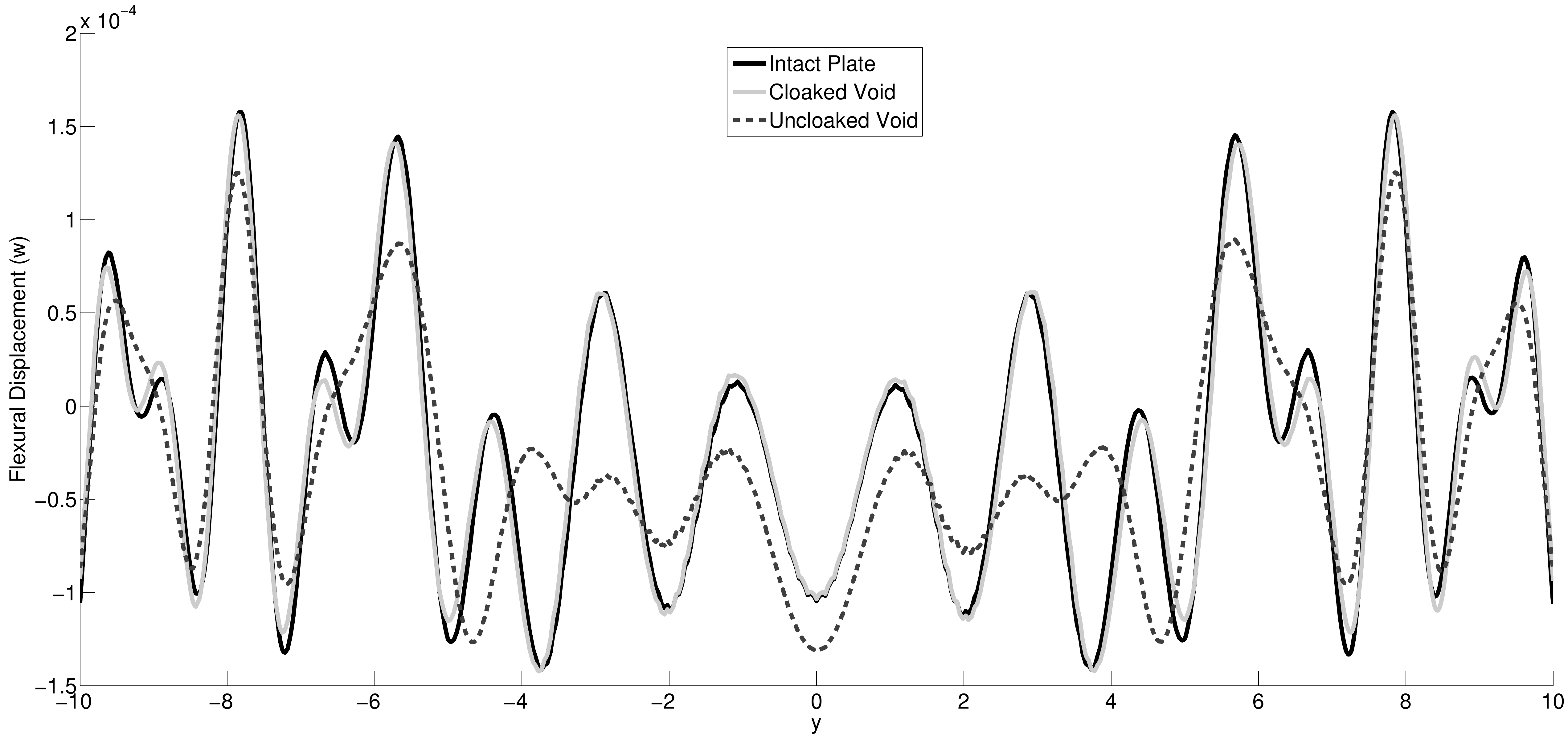}
\setcounter{subfigure}{3}
\caption{\label{fig:Interferometer-Line}}
\end{subfigure}
\caption{\label{fig:Interferometry}
The flexural displacement generated by two coherent point sources in (a) an intact plate, (b) a plate with a square void, and (c) a plate with a cloaked square void.
Figure (d) shows the interference fringes along an observation screen, indicated by dashed lines in figures (a)-(c).
The following parameter values were used: 
$D^{(0)} =19.23$ Nm, $ \Rho =7800$ kg/m$^3$,  $h = 10^{-3}$ m, $a=b=1.0$ m, $\epsilon=10^{-3}$ and $\omega = 157$ rad/s.
}
\end{figure}
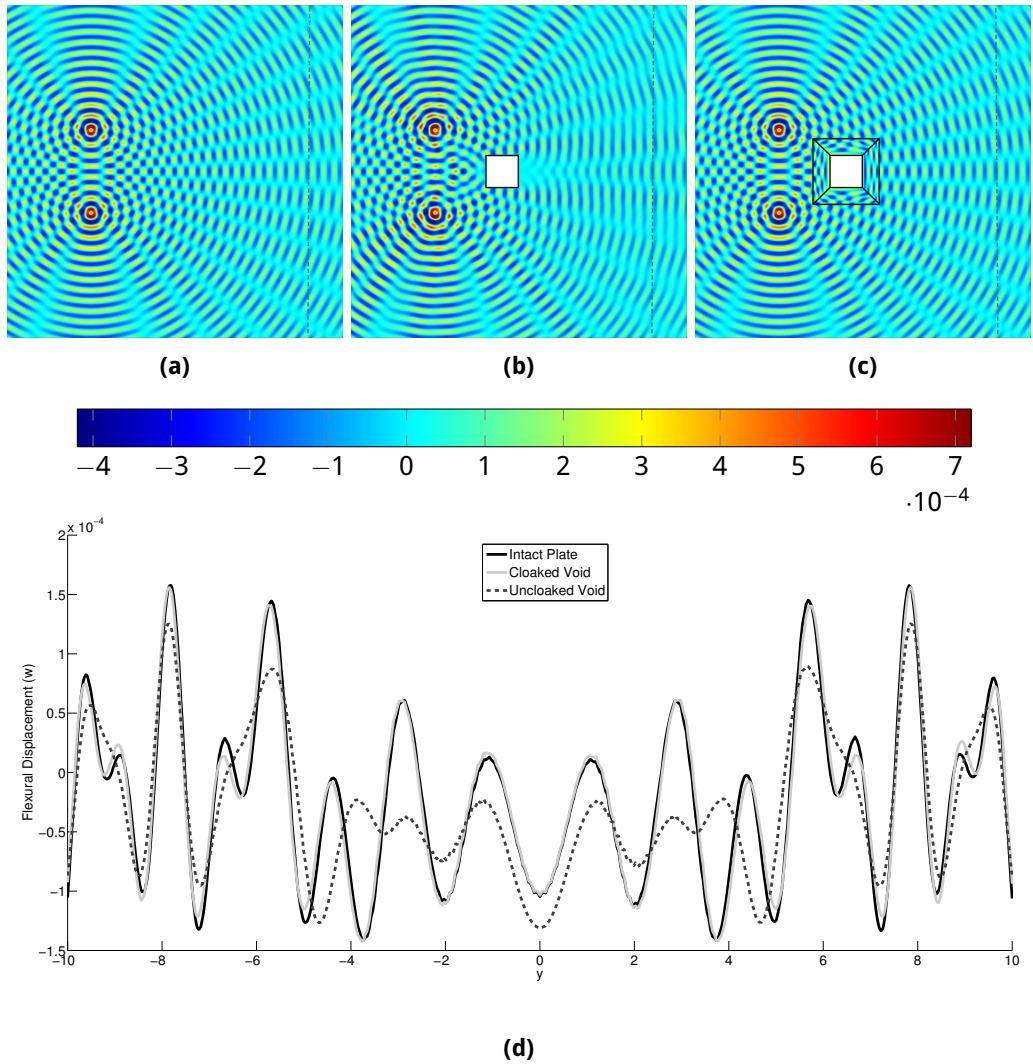

In the present paper the effectiveness of the cloak is examined by a similar method.
Figure~\ref{fig:Interferometry} shows the results of an interferometry simulation were the stability of the interference pattern generated by two coherent cylindrical sources is examined.
Figure~\ref{fig:Interferometry-Intact} shows the interference pattern generated by two coherent cylindrical sources in an infinite homogeneous plate.
Figure~\ref{fig:Interferometry-Uncloaked} shows the perturbation to the interference pattern when a void is introduced, while figure~\ref{fig:Interferometry-Cloaked} shows the interference pattern when the void is coated with a cloak.
The interference patterns seen on the observation screen (dashed
lines in figures~\ref{fig:Interferometry-Intact}--\ref{fig:Interferometry-Cloaked}) for all three cases are shown in figure~\ref{fig:Interferometer-Line}.
It is observed that the interference patterns for the homogeneous plate and cloaked void are virtually identical, whereas the presence of the uncloaked void significantly perturbs the interference pattern.
It is emphasised that this is not a low-frequency response and the dimensions of cloak and inclusion are not small compared to the wavelength.

It is remarked that, conceptually, the interferometry simulation presented in the present paper is equivalent to the classical Young's double-slit interferometer employed in~\cite{colquitt2013}.
In both cases, the interference pattern is very sensitive to perturbations and hence makes a good tool in the measurement of cloaking quality.

\section{Concluding remarks}
\label{sec:conclusion}

A formal framework for transformation elastodynamics as applied to Kirchhoff-Love plates has been developed.
The material properties of the transformed system and the applied pre-stresses and body loads are given explicitly in terms of the deformation gradient in equations~\eqref{eq:mat-params}.
It has been demonstrated that the bi-harmonic equation governing the flexural deformation of a linear homogeneous and isotropic Kirchhoff-Love plate is not invariant under a general coordinate mapping.
Nevertheless, all terms in the transformed equation,~\eqref{eq:transformed-plate-eq} or~\eqref{eq:transformed-plate-eq-index}, can be interpreted in terms of well understood physical quantities within the framework of linear Kirchhoff-Love plate theory.
It is emphasised that the flexural rigidities of the transformed system possess the expected major and minor symmetries, the stresses are symmetric and the density is a scalar function of the Jacobian.

An example transformation for a square push-out cloak complements the formal framework.
It is shown that, for this particular transformation, it is possible to reduce the fully anisotropic plate to a locally orthotropic material.
The analytical work is accompanied by illustrative numerical simulations demonstrating the quality of the cloaking effect, via both numerical measures and interferometery.

The precise physical interpretation of the additional terms in the transformed equation~\eqref{eq:transformed-plate-eq-index} may lead to a refinement of the experimental implementation yielding an improvement of results reported in~\cite{stenger2012} and further developments in the practical implementation of broadband cloaks and metamaterial devices for plates.

\paragraph{Acknowledgements}
{\small
D.J.C., M.G., A.B.M. \& N.V.M. acknowledge the financial support of the European Community's Seven Framework Programme under contract number PIAP-GA-2011-286110-INTERCER2.
M.B. acknowledges the financial support of the European Community's Seven Framework Programme under contract number PIEF-GA-2011-302357-DYNAMETA and of  Regione Autonoma della Sardegna (LR7 2010, grant `M4' CRP-27585).
D.J.C. also acknowledges the financial support of EPSRC in the form of a Doctoral Prize Fellowship and grant EP/J009636/1.
Finally, the authors would like to thank the anonymous referee for their valuable comments.
}

\bibliographystyle{ProcRSocABib}
\bibliography{CloakingPlatesRefs}

\begin{thebibliography}{10}

\bibitem{leonhardt2006}
Leonhardt U. 2006 Optical conformal mapping.
\newblock \emph{Science} \textbf{312}, 1777--1780.

\bibitem{pendry2006}
Pendry JB, Schurig D, Smith DR. 2006 Controlling electromagnetic fields.
\newblock \emph{Science} \textbf{312}, 1780--1782.

\bibitem{schurig2006}
Schurig D, Mock J, Justice B, Cummer S, Pendry J, Starr A, Smith D. 2006
  Metamaterial electromagnetic cloak at microwave frequencies.
\newblock \emph{Science} \textbf{314}, 977--980.

\bibitem{landy2012}
Landy N, Smith DR. 2013 A full-parameter unidirectional metamaterial cloak for
  microwaves.
\newblock \emph{Nature Materials} \textbf{12}, 25--28.

\bibitem{ergin2011}
Ergin T, Fischer J, Wegener M. 2011 Optical phase cloaking of 700 nm light
  waves in the far field by a three-dimensional carpet cloak.
\newblock \emph{Physical Review Letters} \textbf{107}, 173901.

\bibitem{chen2012}
Chen H, Zheng B. 2012 Broadband polygonal invisibility cloak for visible light.
\newblock \emph{Scientific Reports} \textbf{2}, 255.

\bibitem{ward1996}
Ward A, Pendry J. 1996 Refraction and geometry in {M}axwell's equations.
\newblock \emph{Journal of Modern Optics} \textbf{43}, 773--793.

\bibitem{post1962}
Post E. 1962 \emph{Formal Structure of Electromagnetics: General Covariance and
  Electromagnetics}.
\newblock Dover books on physics. Dover Publications.

\bibitem{chen2007}
Chen H, Chan C. 2007 Acoustic cloaking in three dimensions using acoustic
  metamaterials.
\newblock \emph{Applied Physics Letters} \textbf{91}, 183518.

\bibitem{cummer2007}
Cummer SA, Schurig D. 2007 One path to acoustic cloaking.
\newblock \emph{New Journal of Physics} \textbf{9}, 45.

\bibitem{norris2008}
Norris AN. 2008 Acoustic cloaking theory.
\newblock \emph{Proceedings of the Royal Society A: Mathematical, Physical and
  Engineering Science} \textbf{464}, 2411--2434.

\bibitem{chen2010}
Chen H, Chan C. 2010 Acoustic cloaking and transformation acoustics.
\newblock \emph{Journal of Physics D: Applied Physics} \textbf{43}, 113001.

\bibitem{guenneau2012thermo}
Guenneau S, Amra C, Veynante D. 2012 Transformation thermodynamics: cloaking
  and concentrating heat flux.
\newblock \emph{Optics Express} \textbf{20}, 8207--8218.

\bibitem{schittny2013}
Schittny R, Kadic M, Guenneau S, Wegener M. 2013 Experiments on transformation
  thermodynamics: Molding the flow of heat.
\newblock \emph{Physical Review Letters} \textbf{110}, 195901.

\bibitem{han2013}
Han T, Yuan T, Li B, Qiu CW. 2013 Homogeneous thermal cloak with constant
  conductivity and tunable heat localization.
\newblock \emph{Scientific Reports} \textbf{3}, 1593.

\bibitem{parnell2012}
Parnell WJ, Norris AN, Shearer T. 2012 Employing pre-stress to generate finite
  cloaks for antiplane elastic waves.
\newblock \emph{Applied Physics Letters} \textbf{100}, 171907.

\bibitem{parnell2013}
Parnell WJ, Shearer T. 2013 Antiplane elastic wave cloaking using
  metamaterials, homogenization and hyperelasticity.
\newblock \emph{Wave Motion} \textbf{50}, 1140--1152.

\bibitem{colquitt2013}
Colquitt DJ, Jones IS, Movchan NV, Movchan AB, Brun M, McPhedran RC. 2013
  Making waves round a structured cloak: lattices, negative refraction and
  fringes.
\newblock \emph{Proceedings of the Royal Society A: Mathematical, Physical and
  Engineering Science} \textbf{469}, 20130218.

\bibitem{kadic2013}
Kadic M, B{\"u}ckmann T, Schittny R, Wegener M. 2013 Metamaterials beyond
  electromagnetism.
\newblock \emph{Reports on Progress in Physics. Physical Society (Great
  Britain)} \textbf{76}, 126501.

\bibitem{milton2006}
Milton GW, Briane M, Willis JR. 2006 On cloaking for elasticity and physical
  equations with a transformation invariant form.
\newblock \emph{New Journal of Physics} \textbf{8}, 248.

\bibitem{norris2011}
Norris A, Shuvalov A. 2011 Elastic cloaking theory.
\newblock \emph{Wave Motion} \textbf{48}, 525--538.

\bibitem{milton2007}
Milton GW, Willis JR. 2007 On modifications of newton's second law and linear
  continuum elastodynamics.
\newblock \emph{Proceedings of the Royal Society A: Mathematical, Physical and
  Engineering Science} \textbf{463}, 855--880.

\bibitem{brun2009}
Brun M, Guenneau S, Movchan AB. 2009 Achieving control of in-plane elastic
  waves.
\newblock \emph{Applied Physics Letters} \textbf{94}, 061903.

\bibitem{norris2012}
Norris A, Parnell W. 2012 Hyperelastic cloaking theory: transformation
  elasticity with pre-stressed solids.
\newblock \emph{Proceedings of the Royal Society A: Mathematical, Physical and
  Engineering Science} \textbf{468}, 2881--2903.

\bibitem{farhat2009PRL}
Farhat M, Guenneau S, Enoch S. 2009 Ultrabroadband elastic cloaking in thin
  plates.
\newblock \emph{Physical Review Letters} \textbf{103}, 024301.

\bibitem{farhat2009}
Farhat M, Guenneau S, Enoch S, Movchan AB. 2009 Cloaking bending waves
  propagating in thin elastic plates.
\newblock \emph{Physical Review B} \textbf{79}, 033102.

\bibitem{stenger2012}
Stenger N, Wilhelm M, Wegener M. 2012 Experiments on elastic cloaking in thin
  plates.
\newblock \emph{Physical Review Letters} \textbf{108}, 014301.

\bibitem{timoshenko1959}
Timoshenko S, Woinowsky-Krieger S. 1959 \emph{Theory of plates and shells},
  volume~2.
\newblock McGraw-hill New York.

\bibitem{lekhnitskii1968}
Lekhnitskii S, Tsai S, Cheron T. 1968 \emph{Anistropic Plates}.
\newblock New York: Gordon and Breach.

\bibitem{ciarlet1980}
Ciarlet PG. 1980 A justification of the von k{\'a}rm{\'a}n equations.
\newblock \emph{Archive for Rational Mechanics and Analysis} \textbf{73},
  349--389.

\bibitem{blanchard1983}
Blanchard D, Ciarlet P. 1983 A remark on the von karman equations.
\newblock \emph{Computer Methods in Applied Mechanics and Engineering}
  \textbf{37}, 79--92.

\bibitem{kohn2008}
Kohn R, Shen H, Vogelius M, Weinstein M. 2008 Cloaking via change of variables
  in electric impedance tomography.
\newblock \emph{Inverse Problems} \textbf{24}, 015016.

\bibitem{leissa1969}
Leissa AW. 1969 Vibration of plates.
\newblock Technical Report NASA-SP-160, Scientific and Technical Information
  Division, NASA, Washington, DC.

\bibitem{Bigoni2008}
Bigoni D, Gei M, Movchan A. 2008 Dynamics of a prestressed stiff layer on an
  elastic half space: filtering and band gap characteristic of periodic
  structural models derived from long-wave asymptotics.
\newblock \emph{Journal of the Mechanics and Physics of Solids} \textbf{2146},
  2494--2520.

\bibitem{Gei2009}
Gei M, Movchan A, Bigoni D. 2009 Dynamics of a prestressed stiff layer on an
  elastic half space: filtering and band gap characteristic of periodic
  structural models derived from long-wave asymptotics.
\newblock \emph{Journal of Applied Physics} \textbf{105}, 063507.

\bibitem{rahm2008}
Rahm M, Schurig D, Roberts DA, Cummer SA, Smith DR, Pendry JB. 2008 Design of
  electromagnetic cloaks and concentrators using form-invariant coordinate
  transformations of {M}axwell's equations.
\newblock \emph{Photonics and Nanostructures-Fundamentals and Applications}
  \textbf{6}, 87--95.

\bibitem{evans2007}
Evans D, Porter R. 2007 Penetration of flexural waves through a periodically
  constrained thin elastic plate in vacuo and floating on water.
\newblock \emph{Journal of Engineering Mathematics} \textbf{58}, 317--337.

\bibitem{olver2010}
Olver FW, Lozier D, Boisvert R, Clark C, eds. 2010 \emph{NIST handbook of
  mathematical functions}.
\newblock Cambridge University Press.

\end{thebibliography}

\appendix
\section{Material parameters and pre-stress for the cloak}
\label{ap:mat-params}

The material parameters and pre-stresses for the remaining three sides of the cloak are as follows:
\[
\begin{aligned}
D^{(3)}_{1111}  = \alpha_1^2\left(1+\frac{\alpha_2}{x_1}\right)D^{(0)},\;
D^{(3)}_{2222}  = \frac{\alpha_1^2\left(\alpha_2^2x_2^2+x_1^4\right)^2}{\left(x_1+\alpha_2\right)^3x_1^5}D^{(0)},\;
D^{(3)}_{2211}  = \frac{\alpha_1^2\left(\alpha_2^2x_2^2+x_1^4\right)}{\left(x_1+\alpha_2\right)x_1^3}D^{(0)},\\
D^{(3)}_{1212}  = \frac{\alpha_1^2\alpha_2^2x_2^2}{\left(x_1+\alpha_2\right)x_1^3}D^{(0)},\;
D^{(3)}_{1112}  = \alpha_1^2\alpha_2\frac{x_2}{x_1^2}D^{(0)},\;
D^{(3)}_{2212}  = \frac{\alpha_1^2\alpha_2x_2\left(\alpha_2^2x_2^2+x_1^4\right)}{\left(x_1+\alpha_2\right)^2x_1^4}D^{(0)},
\end{aligned}
\]
\[
N^{(3)}_{11} = -\frac{2\alpha_1^2\alpha_2}{x_1^2\left(x_1+\alpha_2\right)}D^{(0)},\;
N^{(3)}_{12} = -\frac{2\alpha_1^2\alpha_2x_2\left(3x_1+2\alpha_2\right)}{\left(x_1+\alpha_2\right)^2x_1^3}D^{(0)},
\]
\[
N^{(3)}_{22} = \frac{2\alpha_1^2\alpha_2\left(x_1^4-8\alpha_2x_2^2x_1-3\alpha_2^2x_2^2\right)}{x_1^4\left(x_1+\alpha_2\right)^3}D^{(0)},
\]
\[
S^{(3)}_1 = 0,\quad S^{(1)}_2 = -\frac{24\alpha_1^2\alpha_2 x_2}{\left(x_1+\alpha_2\right)^3x_1^2}D^{(0)},
\]
\[
\rho^{(3)} = \frac{\Rho\left(x_1+\alpha_2\right)}{\alpha_1^2x_1}.
\]
and
\[
\begin{aligned}
D^{(2,4)}_{1111}  = \frac{\alpha_1^2\left(\alpha_2^2x_1^2+x_2^4\right)^2}{\left(x_2\mp\alpha_2\right)^3x_2^5}D^{(0)},\;
D^{(2,4)}_{2222}  = \alpha_1^2\left(1\mp\frac{\alpha_2}{x_2}\right)D^{(0)},\;
D^{(2,4)}_{2211}  = \frac{\alpha_1^2\left(\alpha_2^2x_1^2+x_2^4\right)}{\left(x_2\mp\alpha_2\right)x_2^3}D^{(0)},\\
D^{(2,4)}_{1212}  = \frac{\alpha_1^2\alpha_2^2x_1^2}{\left(x_2\mp\alpha_2\right)x_2^3}D^{(0)},\;
D^{(2,4)}_{1112}  = \mp\frac{\alpha_1^2\alpha_2x_1\left(\alpha_2^2x_1^2+x_2^4\right)}{\left(x_2\mp\alpha_2\right)^2x_2^4}D^{(0)},\;
D^{(2,4)}_{2212}  = \mp\alpha_1^2\alpha_2\frac{x_1}{x_2^2}D^{(0)},
\end{aligned}
\]
\[
N^{(2,4)}_{11} = \mp\frac{2\alpha_1^2\alpha_2\left(x_2^4\pm8\alpha_2x_1^2x_2-3\alpha_2^2x_1^2\right)}{x_2^4\left(x_2\mp\alpha_2\right)^3}D^{(0)},\;
N^{(2.4)}_{12} = \pm\frac{2\alpha_1^2\alpha_2x_1\left(3x_2\mp2\alpha_2\right)}{\left(x_2\mp\alpha_2\right)^2x_2^3}D^{(0)},
\]
\[
N^{(2,4)}_{22} = \pm\frac{2\alpha_1^2\alpha_2}{x_2^2\left(x_2\mp\alpha_2\right)}D^{(0)},
\]
\[
S^{(2,4)}_1 = \pm\frac{24\alpha_1^2\alpha_2 x_1}{\left(x_2\mp\alpha_2\right)^3x_2^2}D^{(0)}, \quad
S^{(2,4)}_2 = 0,
\]
\[
\rho^{(2,4)} = \frac{\Rho\left(x_2\mp\alpha_2\right)}{\alpha_1^2x_2}.
\]

\end{document}